\documentclass[
aps, prd, reprint,
10pt, notitlepage, a4paper,
floats, floatfix,
amsmath, amssymb, amsfonts,
superscriptaddress,
showpacs, showkeys,
nofootinbib,
longbibliography,
]{revtex4-2}

\usepackage{graphicx} 
\usepackage{grffile}
\usepackage[usenames,dvipsnames]{xcolor}
\usepackage{xspace} 
\usepackage{bm} 
\usepackage{amsmath,amsfonts,amssymb,amsthm}
\usepackage[utf8]{inputenc} 

\usepackage[normalem]{ulem}

\xdefinecolor{mylinkcolor}{rgb}{0,0,0.5}
\usepackage[
	bookmarksnumbered, bookmarksopen, bookmarksopenlevel=2,
	breaklinks=true,
	colorlinks=true, filecolor=mylinkcolor, citecolor=mylinkcolor,
	linkcolor=mylinkcolor, urlcolor=mylinkcolor, menucolor=mylinkcolor,
]{hyperref}
\usepackage{orcidlink}

\def\mr{\mathrm}

\DeclareMathOperator{\Order}{\mathcal{O}}
\def\mF{\mathcal{F}}
\def\pphi{p_\phi}
\def\pphidot{\dot{p}_\phi}
\def\CESA{C_{1\text{ES}^2}}
\def\CESB{C_{2\text{ES}^2}}

\allowdisplaybreaks

\newcommand{\AEI}{\affiliation{Max Planck Institute for Gravitational Physics (Albert Einstein Institute), Am M\"uhlenberg 1, Potsdam 14476, Germany}}
\newcommand{\Maryland}{\affiliation{Department of Physics, University of Maryland, College Park, MD 20742, USA}}

\begin{document}

\title{Radiation-reaction force and multipolar waveforms for eccentric, spin-aligned binaries in the effective-one-body formalism}

\author{Mohammed Khalil\,\orcidlink{0000-0002-6398-4428}}\email{mohammed.khalil@aei.mpg.de}\AEI\Maryland
\author{Alessandra Buonanno\,\orcidlink{0000-0002-5433-1409}}\email{alessandra.buonanno@aei.mpg.de}\AEI\Maryland
\author{Jan Steinhoff\,\orcidlink{0000-0002-1614-0214}}\email{jan.steinhoff@aei.mpg.de}\AEI
\author{Justin Vines\,\orcidlink{0000-0001-6471-5409}}\email{justin.vines@aei.mpg.de}\AEI

\date{\today}

\begin{abstract}
  While most binary inspirals are expected to have circularized before
  they enter the LIGO/Virgo frequency band, a small fraction of those
  binaries could have non-negligible orbital eccentricity depending on
  their formation channel. Hence, it is important to accurately model
  eccentricity effects in waveform models used to detect those
  binaries, infer their properties, and shed light on their
  astrophysical environment. We develop a multipolar
  effective-one-body (EOB) eccentric waveform model for 
  compact binaries whose components have spins aligned or 
anti-aligned with the orbital angular momentum. 
The waveform model contains eccentricity effects in
  the radiation-reaction force and gravitational modes through second
  post-Newtonian (PN) order, including tail effects, and spin-orbit
  and spin-spin couplings. We recast the PN-expanded, eccentric radiation-reaction force 
and modes in factorized form so that the newly derived terms can be directly included 
in the state-of-the-art, quasi-circular--orbit EOB model currently used in LIGO/Virgo analyses 
(i.e., the {\tt SEOBNRv4HM} model).
\end{abstract}

\maketitle

\section{Introduction}

The observation of gravitational waves (GWs) by the LIGO-Virgo
detectors~\cite{LIGOScientific:2020ibl, LIGOScientific:2018mvr} have
corroborated the existence of binary black holes (BBHs)
in our universe.  But how and in which astrophysical environments
these binaries form is not yet fully understood.  However, the masses,
spins (magnitude and orientation), and binary eccentricities inferred from
GWs provide invaluable clues to determine BBH formation
channels~\cite{LIGOScientific:2018jsj,LIGOScientific:2020kqk}.  So far, the
observed GWs are consistent with binary coalescences of negligible
eccentricity, i.e., on quasi-circular orbits~\cite{Salemi:2019owp,
  Romero-Shaw:2019itr,Nitz:2019spj,Romero-Shaw:2020thy}.

In general, binaries are expected to circularize~\cite{Peters:1964zz, Hinder:2007qu} as they approach merger due to the emission of gravitational radiation.
But depending on their astrophysical formation channel, a small fraction of binaries could have non-negligible orbital eccentricity, as they enter the frequency bands of current detectors.
This can occur in dense stellar environments, such as globular clusters or galactic nuclei, where dynamic capture~\cite{Samsing:2017rat,Samsing:2013kua,Samsing:2017xmd,Rodriguez:2017pec,Zevin:2018kzq,Gondan:2020svr} or the Lidov-Kozai mechanism in hierarchical triples~\cite{Antonini:2012ad,Antonini:2015zsa,VanLandingham:2016ccd} can lead to eccentric binary inspirals at close separations.

In particular, Ref.~\cite{Samsing:2017xmd} (and
Ref.~\cite{Rodriguez:2017pec}) showed that $\sim 5\%$ (or $\sim10\%$)
of all mergers in globular clusters enter the LIGO band with
eccentricity $e > 0.1$.  Binaries formed via dynamic capture in
galactic nuclei are expected to have high
eccentricities~\cite{Gondan:2020svr}, with $92\%$ having $e > 0.1$ and
$50\% - 85\%$ having $e > 0.8$ at $10$ Hz.  For a BBH around a
supermassive BH, the Lidov-Kozai mechanism can secularly drive
the BBH to eccentricities near unity for some orientations
\cite{VanLandingham:2016ccd}.  Hence, inferring those eccentricities
from GWs is important for understanding the origin and environment of
BBHs.  Interestingly, Ref.~\cite{Romero-Shaw:2020thy} pointed out that 
GW190521~\cite{Abbott:2020tfl} could be consistent with either an eccentric
nonprecessing or a quasi-circular precessing binary, which illustrates
both the difficulties and prospects of further observations in the
upcoming and future LIGO, Virgo and KAGRA runs~\cite{Abbott:2020qfu}.

While the expected fraction of eccentric GW observations with current
detectors is small, neglecting eccentricity for the parameter
inference can cause significant bias~\cite{Favata:2013rwa}.  This
becomes more relevant for LISA where a large fraction of stellar-mass
binaries is expected to be
eccentric~\cite{Sesana:2010qb,Breivik:2016ddj,Willems:2007xe,Samsing:2018isx,Cardoso:2020iji}.
Hence, it is important to develop accurate waveform models for
eccentric binaries to detect them, infer their properties, and shed
light on their astrophysical environment and formation channels.
Several studies developed post-Newtonian (PN) waveform models for
eccentric orbits, such as
Refs.~\cite{Memmesheimer:2004cv,Boetzel:2017zza,Loutrel:2017fgu,Tanay:2016zog,Yunes:2009yz,Huerta:2014eca,Tiwari:2019jtz,Klein:2018ybm,Damour:2004bz,Konigsdorffer:2006zt,Moore:2016qxz},
or hybrid models that use PN results for the inspiral and
quasi-circular numerical-relativity (NR) simulations near
merger~\cite{Hinder:2017sxy,Huerta:2016rwp,Ramos-Buades:2019uvh}.
Recently, NR simulations for eccentric binaries were reported in
Refs.~\cite{Huerta:2019oxn,Boyle:2019kee,Ramos-Buades:2019uvh}, and
the first NR surrogate model for eccentric BBHs has been developed in
Ref.~\cite{Islam:2021mha}.

The effective-one-body (EOB) formalism~\cite{Buonanno:1998gg,Buonanno:2000ef,Damour:2015isa} improves inspiral-merger-ringdown waveforms by combining information from PN theory, NR simulations, and the strong-field test-body limit.
EOB Hamiltonians have been constructed to include spin~\cite{Damour:2001tu,Damour:2008qf,Barausse:2009xi,Barausse:2011ys,Nagar:2011fx,Balmelli:2013zna,Damour:2014sva,Balmelli:2015zsa,Khalil:2020mmr}, tidal effects~\cite{Damour:2009wj,Bini:2012gu,Steinhoff:2016rfi,Hinderer:2016eia}, information from the small mass-ratio~\cite{Yunes:2010zj,Damour:2009sm,Barausse:2011dq,Akcay:2012ea,Antonelli:2019fmq} and the post-Minkowskian approximations~\cite{Damour:2016gwp,Damour:2017zjx,Antonelli:2019ytb}, and have been refined and calibrated to NR simulations~\cite{Pan:2009wj,Pan:2010hz,Pan:2013rra,Taracchini:2012ig,Taracchini:2013rva,Bohe:2016gbl,Babak:2016tgq,Nagar:2018zoe,Ossokine:2020kjp}. 
While the EOB Hamiltonian is valid for generic orbits, most EOB waveform models use quasi-circular orbit results for the radiation-reaction (RR) force, gravitational waveform modes, and the calibration with NR simulations. 

Recent approaches to extend the EOB formalism to eccentric orbits
  include Ref.~\cite{Bini:2012ji}, which derived the RR force with
  eccentricity up to 2PN order, but without tail effects and for nonspinning
  BHs.  More recently, Ref.~\cite{Hinderer:2017jcs} incorporated
  eccentricity effects in the RR force and in the $(2,2)$ waveform mode 
  through 1.5PN order, including tail effects, using the Keplerian
  parametrization and phase variables that evolve only due to RR. 
  References~\cite{Cao:2017ndf,Liu:2019jpg} extended the quasi-circular
  \texttt{SEOBNRv1}~\cite{Taracchini:2012ig} model to eccentric
  orbits, while Ref.~\cite{Liu:2021pkr} added eccentric corrections 
in the \texttt{SEOBNRv4}~\cite{Bohe:2016gbl,Cotesta:2018fcv} waveform model, 
notably in the $(2, 2), (2, 1), (3, 3),(4,4)$ modes through 2PN order, including spin-orbit (SO) and spin-spin (SS)
  couplings~\footnote{Our results for those modes are mostly in
      agreement with Ref.~\cite{Liu:2021pkr} except for the SO part, where we disagree 
      with their findings (their expressions contain two extra SO terms).}, but not
  tail effects. They employed these eccentric modes to construct a RR force
  for eccentric orbits, which however does not include the Schott terms. As 
  argued in Ref.~\cite{Bini:2012ji} and Sec.~\ref{sec:RRforce} below,
  these Schott terms are necessary for generic orbits to satisfy the flux-balance equations. 
 Furthermore, Refs.~\cite{Chiaramello:2020ehz,Nagar:2021gss} incorporated
 noncircular effects in the \texttt{TEOBResumS\_SM}~\cite{Nagar:2018zoe,Nagar:2020pcj} model 
at leading PN order in the azimuthal component of the RR force, and used a quasi-circular
  2PN-expanded radial RR force without spin or tail effects. They included eccentric corrections 
at leading PN order to all modes $m \neq 0$ up to $\ell=|m|=5$.

In this paper, we develop a multipolar EOB waveform model for
  eccentric binaries with the compact-objects' spins aligned or
  antialigned (henceforth, for short aligned) with the orbital angular
  momentum. We derive the eccentric PN expressions for the RR force (including the Schott terms) and the 
    gravitational modes up to $\ell=|m|=6$, including the $m=0$ mode, through 2PN order, 
    including tail effects, and SO and SS couplings. We recast our
  results for the RR force and modes in a form that can be directly
  incorporated in the state-of-the-art, quasi-circular--orbit EOB
  model currently used in LIGO/Virgo analyses
  (\texttt{SEOBNRv4HM}~\cite{Bohe:2016gbl,Cotesta:2018fcv}).

The paper is structured as follows. In Sec.~\ref{sec:RRforce}, we derive the RR force from the energy and angular momentum fluxes using the balance relations. We use the gauge freedom in the RR force to impose that it reduces to the relation used in \texttt{SEOBNRv4HM} in the quasi-circular--orbit limit.
In Sec.~\ref{sec:initCond}, we obtain initial conditions for eccentric orbits.
In Sec.~\ref{sec:modes}, we calculate all the gravitational waveform modes that contribute up to 2PN order relative to the leading order (LO) of the $(2,2)$ mode, i.e., up to the $\ell=|m|=6$ mode. These higher-order modes are even more important for eccentric orbits than for quasi-circular ones~\cite{Habib:2019cui}.
We conclude in Sec.~\ref{sec:conc} with a discussion of results and potential future work. 
Finally, Appendix~\ref{app:transform} provides the coordinate transformation from harmonic to EOB coordinates, Appendix~\ref{app:SSflux} includes a derivation of the LO spin-squared contribution to the angular momentum flux, Appendix~\ref{app:harmModes} lists the spin contributions to the waveform modes in harmonic coordinates, Appendix~\ref{app:kepler} provides some relations for dynamic quantities in the Keplerian parametrization, and Appendix~\ref{app:tort} includes the transformation to tortoise coordinates.
We provide our results for the RR force and waveform modes as \textit{Mathematica} files in the Supplemental Material~\cite{ancmaterial}.

\subsection*{Notation}
We use the metric signature $(-,+,+,+)$, and use units in which $c = G =1$, but write $c$ explicitly in PN expansions.

We consider an aligned-spin binary with masses $m_1$ and $m_2$, with $m_1 \geq m_2$, and we define the following constants:
\begin{gather}
M= m_1 + m_2, \quad \mu = \frac{m_1m_2}{M}, \quad \nu = \frac{\mu}{M},  \nonumber\\ 
\delta =\frac{m_1 - m_2}{M},  \quad
X_1 = \frac{m_1}{M}, \quad 
X_2 = \frac{m_2}{M}.
\end{gather}

In the binary's center of mass, we introduce the canonical phase-space variables $(R,\phi,P_R,P_\phi)$, where $R$ is the separation, $\phi$ the azimuthal angle, $P_R$ the radial momentum, and $P_\phi$ the angular momentum. The total relative momentum $P$ is given by $P^2 = P_R^2 + P_\phi^2/R^2$.
We use the rescaled dimensionless variables
\begin{gather}
r=\frac{R}{M}, \quad t= \frac{T}{M}, \quad p_r = \frac{P_r}{\mu}, \quad 
\pphi=\frac{P_\phi}{M\mu}, \nonumber\\
\hat{H}=\frac{H}{\mu}, \quad \hat{S}_i = \frac{S_i}{M\mu}, \quad \chi_i = \frac{S_i}{m_i^2},
\end{gather}
where the dimensionless quantities are denoted with either a hat or a lowercase letter.

The energy and angular momentum fluxes far away from the binary are denoted by $\Phi_E$ and $\Phi_J$ respectively, and scale as follows:
\begin{equation}
\Phi_E = G c^5 \tilde{\Phi}_E, \quad \Phi_J = c^5 \frac{\tilde{\Phi}_J}{M},
\end{equation}
where quantities with a tilde are the physical dimensionful fluxes. The components of the RR force are denoted by $\mF_r$ and $\mF_\phi$, and are scaled similarly to $\Phi_E$ and $\Phi_J$, respectively.

\section{Radiation reaction force}
\label{sec:RRforce}
The RR force accounts for the energy and angular momentum losses by the system, and is added to the right-hand side of the Hamilton equations of motion (EOMs) such that
\begin{align}
\label{EOMs}
\dot{r} = \frac{\partial \hat{H}}{\partial p_r}, \qquad \dot{p}_r = - \frac{\partial \hat{H}}{\partial r} + \mathcal{F}_r,  \nonumber\\
\dot{\phi} = \frac{\partial \hat{H}}{\partial \pphi}, \qquad  \pphidot = - \frac{\partial \hat{H}}{\partial \phi} + \mathcal{F}_\phi,
\end{align}
where the leading order of $\mathcal{F}_{r,\phi}$ is of order $1/c^5$ (2.5 PN).
From the EOMs, with $\partial H / \partial \phi= 0$, the time derivatives of energy and angular momentum are given by
\begin{align}
&\dot{E}_\text{system} = \frac{d\hat{H}}{dt} = \dot{r} \mathcal{F}_r + \dot{\phi} \mathcal{F}_\phi, \nonumber\\
&\dot{J}_\text{system} = \frac{d\pphi}{dt} = \mathcal{F}_\phi.
\end{align}
The energy and angular momentum lost by the system are not equal to the energy and angular momentum fluxes, $\Phi_E$ and $\Phi_J$, because of additional contributions to $E$ and $J$ due to interactions with the radiation field. The balance equations are modified by \emph{Schott} terms, as in electrodynamics, that appear as total time derivatives in the balance equations~\cite{Bini:2012ji}
\begin{align}
&\dot{E}_\text{system} + \dot{E}_\text{Schott} + \Phi_E = 0, \nonumber\\
&\dot{J}_\text{system} + \dot{J}_\text{Schott} + \Phi_J = 0.
\end{align}
Substituting the expressions for the energy and angular momentum losses, we obtain
\begin{align}
\label{baleqns}
&\dot{r} \mathcal{F}_r + \dot{\phi} \mathcal{F}_\phi + \dot{E}_\text{Schott} + \Phi_E = 0, \nonumber\\
&\mathcal{F}_\phi + \dot{J}_\text{Schott} + \Phi_J = 0.
\end{align}
The energy and angular momentum fluxes are gauge-independent, but the RR force and Schott terms are gauge-dependent. This coordinate gauge freedom in the RR force was discussed by Iyer and Will in Refs.~\cite{Iyer:1993xi, Iyer:1995rn}, and by Gopakumar et. al. in Ref.~\cite{Gopakumar:1997ng}. Bini and Damour showed in Ref.~\cite{Bini:2012ji} how the gauge freedom in $\mF$ is related to the freedom in defining the Schott terms.

Note that while we only consider aligned spins in this paper, an extension to precessing spins is straightforward; the RR force $\bm{\mF}$ is added to the EOM for the total momentum $\bm{p}$ and a RR contribution is added to the spin evolution equations, such that
\begin{gather}
\frac{d\bm{r}}{dt} = \frac{\partial H}{\partial \bm{p}}, \qquad 
\frac{d\bm{p}}{dt} = -\frac{\partial H}{\partial \bm{r}} + \bm{\mF}, \nonumber\\
\frac{d\bm{S}_{\mr i}}{dt} =  \frac{\partial H}{\partial \bm{S}_{\mr i}} \times \bm{S}_{\mr i} + \dot{\bm{S}}_{\mr i}^\text{RR}. 
\end{gather}
The balance equations are then given by
\begin{align}
&\dot{E}_\text{system} + \dot{E}_\text{Schott} + \Phi_E = 0 \nonumber\\
&\dot{\bm{J}}_\text{system} + \dot{\bm{J}}_\text{Schott} + \bm{\Phi}_J = 0,
\end{align}
with
\begin{align}
\dot{E}_\text{system} &= \dot{\bm{r}}\cdot \bm{\mF}, \nonumber\\
\dot{\bm{J}}_\text{system} &=\bm{r} \times \bm{\mF} + \dot{\bm{S}}_1^\text{RR} + \dot{\bm{S}}_2^\text{RR}.
\end{align}
See, e.g., Refs.~\cite{Zeng:2007bq,Wang:2007ntb} for more details.

\subsection{Summary of the approach used in this paper for the RR force}

The aim of this paper is to extend the quasi-circular RR force and gravitational modes employed in the \texttt{SEOBNRv4HM} waveform model to eccentric orbits.
The Hamilton equations that describe the dynamics of the \texttt{SEOBNRv4HM} model use the following relations between the RR force and the energy flux for quasi-circular orbits, which are based on results from Refs.~\cite{Buonanno:2000ef, Buonanno:2005xu},
\begin{align}
\label{RRFcirc}
\mF_\phi^\text{qc} &= - \frac{\Phi_E^\text{qc}}{\Omega}, \nonumber\\ 
\mF_r^\text{qc} &= \mF_\phi^\text{qc} \frac{p_r}{\pphi} = -  \frac{\Phi_E^\text{qc} p_r}{\Omega \pphi},
\end{align}
with $\Omega$ being the (angular) orbital frequency. 
However, these two relations are only valid for quasi-circular orbits and are not consistent for generic orbits, since they use the circular-orbit relation $\Phi_E^\text{qc} = \Omega \Phi_J^\text{qc}$ and do not include the Schott terms.

Hence, the approach we use to obtain the RR force is to write a generic ansatz with unknown coefficients for the Schott terms, and calculate the RR force from the fluxes using the balance equations
\begin{align}
\label{balance}
\mF_\phi &= - \Phi_J - \dot{J}_\text{Schott}, \nonumber\\
\dot{r} \mF_r &= - \Phi_E + \dot{\phi} \Phi_J - \dot{E}_\text{Schott} + \dot{\phi} \dot{J}_\text{Schott}.
\end{align}
Then, we specify the free unknown coefficients in the Schott terms such that the force reduces to the conditions in Eq.~\eqref{RRFcirc} in the limit of quasi-circular orbits, i.e.,
\begin{align}
\label{Fconds}
\mF_\phi &= -\Phi_J + \Order(\dot{p}_r) + \Order(p_r^2), \nonumber\\ 
\frac{\mF_r \pphi}{\mF_\phi p_r} &= 1 + \Order(p_r^2),
\end{align}
since both $p_r$ and $\dot{p}_r$ are zero for circular orbits.
Finally, we factorize the RR force into the quasi-circular part used in \texttt{SEOBNRv4HM} times eccentric corrections
\begin{equation}
\label{Ffact}
\mF_r = \mF_r^\text{qc} \mF_r^\text{ecc}, \qquad \mF_\phi = \mF_\phi^\text{qc} \mF_\phi^\text{ecc},
\end{equation}
where the quasi-circular parts are given by Eq.~\eqref{RRFcirc}, and the eccentric corrections scale as $\mF_i^\text{nc} \sim 1 + \dot{p}_r + p_r^2 + \dots$. In the following subsections, we provide the details of these steps.

\subsection{EOB Hamiltonian and angular momentum}
The EOB Hamiltonian is calculated from an effective Hamiltonian $H_\text{eff}$ via the energy map
\begin{equation}
H_\text{EOB} = M \sqrt{1 + 2\nu \left(\frac{H_\text{eff}}{\mu} - 1\right)},
\end{equation}
with $H_\text{eff}$ given in Refs.~\cite{Barausse:2009xi,Barausse:2011ys,Khalil:2020mmr}.
When calculating the RR force to 2PN, we only need to work with the PN expansion of the EOB Hamiltonian.
The nonspinning part to 2PN order is given by
\begin{align}
\hat{H}_\text{EOB}^0 &= \frac{c^2}{\nu} + \frac{p^2}{2} -\frac{1}{r} + \frac{1}{c^2}\bigg[\frac{(\nu -1) p^2}{2 r}-\frac{1+\nu}{8} p^4  -\frac{p_r^2}{r} \nonumber\\
&\quad-\frac{1+\nu}{2 r^2} \bigg] 
+ \frac{1}{c^4}\bigg[
\frac{1+\nu + \nu ^2}{16} p^6 +\frac{(1+2 \nu)p_r^2}{r^2} \nonumber\\
&\quad -\frac{\left(1+\nu-3 \nu ^2\right) p^2}{4 r^2}+\frac{\left(1+\nu-3 \nu ^2\right) p^4}{8 r} \nonumber\\
&\quad+ \frac{(1+\nu) p^2p_r^2}{2 r} -\frac{1-\nu+\nu ^2}{2 r^3} \bigg],
\end{align} 
the LO (1.5PN) spin-orbit part
\begin{equation}
\hat{H}_\text{EOB}^\text{SO} = \frac{\pphi}{2 c^3 r^3} \left[\chi_1 \left(2 + 2 \delta -\nu\right) + \chi_2 \left(2 - 2 \delta -\nu\right)\right],
\end{equation}
and the LO (2PN) spin-spin part
\begin{align}
\hat{H}_\text{EOB}^\text{SS} &= \frac{1}{2 c^4 r^3} \Bigg\lbrace
\chi _1^2 \left[X_1^4 \left(1-\frac{\pphi^2}{r}+r p_r^2\right)-\CESA X_1^2\right] \nonumber\\
&\quad +\chi _2^2 \left[X_2^4 \left(1-\frac{\pphi^2}{r}+r p_r^2\right)-\CESB X_2^2\right] \nonumber\\
&\quad
 + 2 \chi _1 \chi _2 \left[(\nu -1) \nu -\frac{\nu ^2 \pphi^2}{r}+\nu ^2 r p_r^2\right] \Bigg\rbrace,
\end{align}
where $C_{i\text{ES}^2}$ are the spin quadrupole constants, which equal one for BHs.

The orbital frequency expanded to 2PN is given by
\begin{align}
\Omega &\equiv \dot{\phi} = \frac{\partial \hat{H}_\text{EOB}}{\partial \pphi} \nonumber\\
&= \frac{\pphi}{r^2}+\frac{\pphi}{c^2}\left[\frac{\nu -1}{r^3}-\frac{(\nu +1) p^2}{2 r^2}\right] \nonumber\\
&\quad +\frac{1}{2c^3r^3} \left[\chi_1 (2 + 2 \delta -\nu) + \chi_1 (2 - 2 \delta -\nu)\right] \nonumber\\
&\quad 
+\frac{\pphi}{c^4} \bigg[\frac{3 \left(\nu ^2+\nu +1\right) p^4}{8 r^2}+\frac{\left(-3 \nu ^2+\nu +1\right) p^2}{2 r^3} \nonumber\\
&\quad\qquad +\frac{(\nu +1) p_r^2}{r^3}-\frac{-3 \nu ^2+\nu +1}{2 r^4}\bigg] \nonumber\\
&\quad
-\frac{\pphi}{c^4r^4} \left(2 \nu ^2 \chi _1 \chi _2 + \chi _1^2 X_1^4 + \chi _2^2 X_2^4\right)
\end{align}

From the EOM $\dot{p}_r = -\partial\hat{H}/\partial r$, we can obtain an expression for $\pphi(r,\dot{p}_r,p_r)$
\begin{align}
\label{Lprd}
\frac{\pphi^2}{r} &= 1 + r^2 \dot{p}_r 
+\frac{1}{2c^2 r} \Big[6+(\nu +1) r^4 \dot{p}_r^2-(\nu -5) r^2 \dot{p}_r \nonumber\\
&\quad\qquad 
+r p_r^2 \left((\nu +1) r^2 \dot{p}_r+4\right)\Big] \nonumber\\
&\quad - \frac{3 \sqrt{r^3 \dot{p}_r+r}}{2c^3 r^2} \left[\chi _1 (2 + 2 \delta -\nu) + \chi _2 (2 - 2 \delta -\nu)\right] \nonumber\\
&\quad + \frac{1}{8c^4 r^2} \Big[
\left(\nu ^2+5 \nu +1\right) r^6 \dot{p}_r^3-\left(\nu ^2-\nu +1\right) r^4 p_r^4 \dot{p}_r \nonumber\\
&\quad\qquad+2 (5 \nu +8) r^4 \dot{p}_r^2-\left(\nu ^2+7 \nu -63\right) r^2 \dot{p}_r \nonumber\\
&\quad\qquad  -24 (\nu -3)+ 2 \left(8-24 \nu +3 \nu  r^4 \dot{p}_r^2\right)r p_r^2 \nonumber\\
&\quad\qquad +2\left(\nu ^2+\nu +3\right) p_r^2 r^3 \dot{p}_r
\Big] \nonumber\\
&\quad + \frac{1}{2c^4 r^2} \bigg\lbrace\!
\chi _1^2 \left[3 \CESA X_1^2 \!+\! X_1^4 \left(1 + 4 r^2 \dot{p}_r-2 r p_r^2\right)\right] \nonumber\\
&\quad\quad+ \chi _2^2 \left[3 \CESB X_2^2+X_2^4 \left(1 + 4 r^2 \dot{p}_r-2 r p_r^2\right)\right] \nonumber\\
&\quad\quad+ \nu\chi _1 \chi _2 \left[2   \left(\nu +4 \nu  r^2 \dot{p}_r+3\right)-4 \nu r p_r^2\right]
\bigg\rbrace,
\end{align}
which we use to express the noncircular part of the RR force and modes in terms of $p_r$ and $\dot{p}_r$.
It will also be useful below, when taking the circular-orbit limit, to have an expression for $\pphi$ as a function of $r$ for circular orbits. Setting $\dot{p}_r = 0 = p_r$ in the previous equation yields
\begin{align}
\label{Lcirc}
\frac{\pphi^2}{r} &\overset{\text{circ}}{=} 1 + \frac{3}{c^2r} + \frac{3(3-\nu)}{c^4r^2}\nonumber\\
&\quad - \frac{3}{2c^3r^{3/2}} \left[\chi_1  \left(-2 \delta +\nu -2\right) + \chi_2  \left(2 \delta +\nu -2\right)\right] \nonumber\\
&\quad + \frac{1}{2c^4r^2} \bigg[\chi _1^2 \left(3 \CESA X_1^2 + X_1^4 \right)
+  2\nu\chi _1 \chi _2  \left(3 + \nu\right) \nonumber\\
&\quad\qquad + \chi _2^2 \left(3 \CESB X_2^2 + X_2^4 \right) 
 \bigg].
\end{align}

\subsection{Energy and angular momentum fluxes}
The energy and angular momentum fluxes for nonspinning binaries were derived to 3PN order in harmonic and Arnowitt-Deser-Misner (ADM) coordinates in Refs.~\cite{Arun:2007sg,Arun:2007rg,Arun:2009mc}.
The 2PN instantaneous part of the fluxes for nonspinning bodies is given in EOB coordinates in Appendix A of Ref.~\cite{Bini:2012ji}. The leading order reads
\begin{align}
\Phi_E^\text{inst} &= \frac{8 \nu^2}{15r^4} \left(12 p^2 - 11 p_r^2\right) + \Order\left(\frac{1}{c^2}\right), \nonumber\\
\Phi_J^\text{inst} &= \frac{8\nu^2}{5r^3} \pphi \left(2 p^2 - 3 p_r^2 + \frac{2}{r}\right) + \Order\left(\frac{1}{c^2}\right).
\end{align}

The hereditary contributions to the fluxes can be expressed as an infinite sum over Bessel functions~\cite{Hinderer:2017jcs,Arun:2007rg} that can be evaluated numerically, or resummed analytically~\cite{Loutrel:2016cdw,Tanay:2016zog}.
Here, we follow the method from Ref.~\cite{Hinderer:2017jcs} to obtain the LO tail part (1.5 PN) of the orbit-averaged fluxes in an eccentricity expansion and we extend their derivation to $\Order(e^6)$, which yields\footnote{Calculating the tail contribution to the fluxes is similar to that for the waveform modes (see Sec.~\ref{sec:modesTail}) except for using the integrals~\cite{Blanchet:2013haa}
\begin{align}
\Phi_E^\text{tail} &= \frac{4}{5} I_{ij}^{(3)} \int_{0}^{\infty} d\tau \, I_{ij}^{(5)}(t-\tau) \ln\left(\frac{\tau}{b}\right), \nonumber\\
\Phi_J^\text{tail} &= \frac{4}{5} \epsilon_{zij}\bigg[ I_{il}^{(2)} \int_{0}^{\infty} d\tau \, I_{jl}^{(5)}(t-\tau) \ln\left(\frac{\tau}{b}\right) \nonumber\\
&\quad\qquad +I_{jl}^{(3)} \int_{0}^{\infty} d\tau \, I_{il}^{(4)}(t-\tau) \ln\left(\frac{\tau}{b}\right)\bigg],
\end{align}
where $I_{ij}$ is the mass quadrupole moment, and $b = 2r_0 \mathrm{e}^{-11/12}$ with $r_0$ a gauge parameter. 
}
\begin{align}
\label{avgFluxTail}
\left\langle \Phi_E^\text{tail} \right\rangle &= \frac{128\pi \nu^2 }{5c^3} x^{13/2} \bigg[1 + \frac{2335}{192} e^2 + \frac{42955}{768} e^4 \nonumber\\
&\quad\qquad + \frac{6204647}{36864} e^6 + \Order(e^8)\bigg], \nonumber\\
\left\langle \Phi_J^\text{tail} \right\rangle &= \frac{128\pi \nu^2 }{5c^3} x^5 \bigg[1 + \frac{209}{32} e^2 + \frac{2415}{128} e^4 + \frac{730751 }{18432}e^6 \nonumber\\
&\quad\qquad + \Order(e^8)\bigg],
\end{align}
where $x \equiv \Omega^{2/3}$.
The eccentricity $e$ in these equations is defined using the Keplerian parametrization, which is given by 
\begin{equation}
r = \frac{1}{u_p (1 + e \cos \chi)}\,,
\end{equation} 
where $u_p$ is the inverse semilatus rectum and $\chi$ is the relativistic anomaly.

Since we are not using the adiabatic approximation and are not working with orbit-averaged fluxes, we can obtain an approximate expression for the tail contribution to the fluxes by writing an ansatz in terms of $(r,p_r,\pphi)$ in a $p_r$ expansion of the form
\begin{align}
\Phi_E^\text{tail} &= \frac{128\pi\nu^2 \pphi}{5c^3r^4}  \left[\frac{1}{r^3} + c_1 \frac{p_r^2}{r^2}  + c_2 \frac{p_r^4}{r} + c_3 p_r^6  + \Order(p_r^8)\right], \nonumber\\
\Phi_J^\text{tail} &= \frac{128\pi  \nu^2}{5c^3r^2}  \left[\frac{1}{r^3} + c_4 \frac{p_r^2}{r^2}+ c_5 \frac{p_r^4}{r} + c_6 p_r^6 + \Order(p_r^8)\right],
\end{align}
calculate the average of that ansatz in terms of $(e,x)$ (see Appendix~\ref{app:kepler}), and then match it to the average flux in Eq.~\eqref{avgFluxTail} to determine the unknowns $c_n$. This yields
\begin{align}
\Phi_E^\text{tail} &= \frac{128\pi\nu^2 \pphi}{5c^3r^4}  \left[\frac{1}{r^3} +\frac{415p_r^2}{96r^2}  +\frac{5p_r^4}{288r}  -\frac{73p_r^6}{11520}  + \Order(p_r^8)\right], \nonumber\\
\Phi_J^\text{tail} &= \frac{128}{5c^3r^2} \pi  \nu^2 \left[\frac{1}{r^3} + \frac{49p_r^2}{16r^2} -\frac{49p_r^6}{5760} + \Order(p_r^8)\right].
\end{align}

The LO (1.5PN) SO fluxes for generic orbits and generic spins were derived in Refs.~\cite{Kidder:1995zr,Zeng:2007bq}. (The next-to-leading-order (NLO) SO energy flux was derived in Ref.~\cite{Bohe:2013cla}.)
It should be noted that Ref.~\cite{Kidder:1995zr} used the Tulczyjew-Dixon (covariant) spin supplementary condition (SSC)~\cite{Dixon:1979,Steinhoff:2014kwa,Tulczyjew:1959,fokker1929relativiteitstheorie}, while Ref.~\cite{Zeng:2007bq} used the Newton-Wigner (NW), or canonical, SSC~\cite{pryce1948mass,newton1949localized}.
In this paper, we use the NW SSC since we are working in a canonical Hamiltonian formulation of the spinning two-body dynamics~\cite{Vines:2016unv,Barausse:2009aa}.
Changing the velocities in Eq. (17) of Ref.~\cite{Zeng:2007bq} to momenta, which involves spin-orbit terms, the aligned-spin fluxes reduce to
\begin{align}
\Phi_E^\text{SO} &= \frac{4\nu^2 \pphi}{15c^3r^6} \bigg\lbrace 
\! \chi_1 \bigg[p^2 (36 \nu  -37 -37 \delta )+\frac{4 (9+9 \delta -4 \nu)}{r}  \nonumber\\
&\quad
+9 p_r^2 (3 + 3 \delta -2 \nu)\bigg] + 1\leftrightarrow 2\bigg\rbrace, \nonumber\\
\Phi_J^\text{SO} &= \frac{4\nu^2}{15 c^3r^3} \bigg\lbrace 
\chi_1 \bigg[p^4 (8 \nu-9 \delta -9)  +\frac{9 + 9 \delta -4 \nu}{r^2}  \nonumber\\
&\quad +(9 + 9 \delta -24 \nu) p^2 p_r^2  - (17 + 17 \delta +6 \nu)\frac{ p_r^2}{r} \nonumber\\
&\quad 
+15 \nu  p_r^4 + \left(11 + 11 \delta +10 \nu\right)\frac{p^2}{r}
\bigg] + 1\leftrightarrow 2
\bigg\rbrace.
\end{align}

For the LO (2PN) SS contributions, the LO spin$_1$-spin$_2$ energy and angular momentum fluxes in harmonic coordinates were derived in Refs.~\cite{Kidder:1995zr, Wang:2007ntb}, while the spin-squared energy flux was derived in Refs.~\cite{Maia:2017yok,Bohe:2015ana}, and we obtain in Appendix~\ref{app:SSflux} the spin-squared angular momentum flux.
Transforming from harmonic to EOB coordinates, using the transformations in Appendix~\ref{app:transform}, we get the following SS contributions to the fluxes for aligned-spins:
\begin{align}
\Phi_E^\text{SS} &= \frac{\nu^2}{15c^4r^6} \Big\lbrace
\chi_1^2 \Big[\CESA (\delta -2 \nu +1) \left(144 p^2-156 p_r^2\right) \nonumber\\
&\quad\quad  + 3 \left(96 \nu\delta -47 \delta -96 \nu ^2+190 \nu -47\right) p^2 \nonumber\\
&\quad\quad + \left(149 \delta -280 \nu \delta +280 \nu ^2-578 \nu +149\right) p_r^2 \Big] \nonumber\\
&\quad + \nu \chi _1 \chi _2 \left[10 (28 \nu -33) p_r^2-6  (48 \nu -47) p^2\right] \nonumber\\
&\quad
+ 1\leftrightarrow 2 \Big\rbrace. \nonumber\\
\Phi_J^\text{SS} &= \frac{\nu^2 \pphi}{5c^4r^5} \bigg\lbrace
\chi_1^2 \bigg[\left(32 \nu\delta -15\delta-32 \nu ^2+62 \nu -15\right) \frac{1}{r} \nonumber\\
&\quad\quad +  \left(\delta-2 \delta  \nu +2 \nu ^2-4 \nu +1\right) \left( 22 p_r^2 - 20 p^2\right) \nonumber\\
&\quad\quad + \CESA (\delta -2 \nu +1) \left(12 p^2-30 p_r^2+\frac{24}{r}\right)  \bigg]\nonumber\\
&\quad  + \nu \chi_1 \chi_2 \bigg[ \frac{2 (23-16 \nu ) }{r}-8   (5 \nu -3) p^2 \nonumber\\
&\quad\quad +4   (11 \nu -15) p_r^2
\bigg]   + 1\leftrightarrow 2
\bigg\rbrace.
\end{align}

The total 2PN energy and angular momentum fluxes are the sum of all the above contributions, i.e.,
\begin{align}
\Phi_E = \Phi_E^\text{inst} + \Phi_E^\text{tail} + \Phi_E^\text{SO} + \Phi_E^\text{SS}, \nonumber\\
\Phi_J = \Phi_J^\text{inst} + \Phi_J^\text{tail} + \Phi_J^\text{SO} + \Phi_J^\text{SS}.
\end{align}

\subsection{Ansatz for the Schott terms}
As an ansatz for the Schott terms $E_\text{Schott}$ and $J_\text{Schott}$, we consider
\begin{align}
\label{SchottInst}
J_\text{Schott}^\text{inst} &= \frac{\nu^2 p_r \pphi}{r^2} \bigg[\alpha_1
+\frac{1}{c^2} \left(\alpha _2 p_r^2+\alpha _3 p^2+ \frac{\alpha_4}{r} \right)  \nonumber\\
&\quad +\frac{1}{c^4} \bigg(\alpha _5 p_r^4+\alpha _6 p^2 p_r^2 +\alpha _7 \frac{p_r^2}{r} + \alpha _8 p^4 +\alpha _9 \frac{p^2}{r} \nonumber\\
&\quad\qquad  +\frac{\alpha _{10}}{r^2}\bigg)
\bigg],\nonumber \\
E_\text{Schott}^\text{inst} &= \frac{\nu^2 p_r}{r^2} \bigg[\beta_1 p_r^2 + \beta_2 p^2 + \frac{\beta_3}{r} 
+\frac{1}{c^2} \bigg(\beta _4 p_r^4 +\beta _5 p^2 p_r^2\nonumber\\
&\quad +\beta _6 \frac{p_r^2}{r}+\beta _7 p^4+\beta _8 \frac{p^2}{r}+ \frac{\beta _9}{r^2}  \bigg) \nonumber\\
&\quad +\frac{1}{c^4} \bigg(
\beta _{10} p_r^6 +\beta _{11} p^2 p_r^4 +\beta _{12} \frac{p_r^4}{r} +\beta _{13} p^4 p_r^2   \nonumber\\
&\quad\qquad  +\beta _{14} \frac{p^2p_r^2}{r}  + \beta _{15} \frac{p_r^2}{r^2} +\beta _{16} p^6+\beta _{17} \frac{p^4}{r}  \nonumber\\
&\quad\qquad +\beta _{18} \frac{p^2}{r^2}  + \frac{\beta _{19}}{r^3}
\bigg)\bigg].
\end{align}
Note that this ansatz for $E_\text{Schott}$ is more general than the one used in Eq.~(4.4) of Ref.~\cite{Bini:2012ji}, since we found that such an ansatz is needed for the RR force to satisfy the conditions in Eq.~\eqref{RRFcirc}.

For the LO tail, we use the ansatz
\begin{align}
J_\text{Schott}^\text{tail} &= \frac{\pi \nu^2 p_r}{c^3r^2} \left(\lambda _1 p_r^2 + \lambda _2 p^2 +\frac{\lambda _3}{r} \right), \nonumber\\
E_\text{Schott}^\text{tail} &= \frac{\pi \nu^2 \pphi p_r}{c^3r^4} \left(\lambda _4 p_r^2 + \lambda _5 p^2+\frac{\lambda _6}{r}\right),
\end{align}
while for the LO SO part,
\begin{align}
J_\text{Schott}^\text{SO} &= \frac{\nu^2 p_r }{c^3r^2} \bigg[
\chi_1 \left(\sigma_1 p_r^2 + \sigma_2 p^2 + \frac{\sigma_3}{r}\right) \nonumber\\
&\quad\quad +\chi_2 \left(\sigma_4 p_r^2 + \sigma_5 p^2 + \frac{\sigma_6}{r} \right)
\bigg], \nonumber\\
E_\text{Schott}^\text{SO} &= \frac{\nu^2 p_r \pphi}{c^3r^4} \bigg[
\chi_1 \left(\sigma_7 p_r^2 + \sigma_8 p^2 + \frac{\sigma_9}{r}\right) \nonumber\\
&\quad\quad +\chi_2 \left(\sigma_{10} p_r^2 + \sigma_{11} p^2 + \frac{\sigma_{12}}{r}\right)
\bigg],
\end{align}
and for the SS part,
\begin{align}
\label{SchottSS}
J_\text{Schott}^\text{SS} &= \frac{\nu^2 \pphi p_r}{c^4r^4}  \bigg[\chi_1^2 \left(\zeta _1 + \CESA \zeta _2\right)  \nonumber\\
&\quad + \chi_1 \chi_2  \zeta_3  + \chi_2^2\left(\zeta _4 + \CESB \zeta _5\right)\bigg], \nonumber\\
E_\text{Schott}^\text{SS} &= \frac{\nu^2 p_r}{c^4r^4} \bigg[
\chi_1^2 \left(\zeta _6 p_r^2+\zeta _7 p^2+\frac{\zeta _8}{r} \right)\nonumber\\
&\quad + \chi_1^2 \CESA\left(\zeta _9 p_r^2+\zeta _{10} p^2+ \frac{\zeta _{11}}{r}\right) \nonumber\\
&\quad + \chi_1 \chi_2 \left(\zeta _{12} p_r^2+\zeta _{13} p^2+ \frac{\zeta _{14}}{r}\right) \nonumber\\
&\quad + \chi_2^2 \left(\zeta _{15} p_r^2+\zeta _{16} p^2+ \frac{\zeta _{17}}{r}\right)\nonumber\\
&\quad + \chi_2^2 \CESB \left(\zeta _{18} p_r^2+\zeta _{19} p^2+ \frac{\zeta _{20}}{r} \right)\!
\bigg].
\end{align}

The total energy and angular momentum Schott terms are the sum of the above contributions, i.e.
\begin{align}
J_\text{Schott} &= J_\text{Schott}^\text{inst} + J_\text{Schott}^\text{tail} + J_\text{Schott}^\text{SO} + J_\text{Schott}^\text{SS}, \nonumber\\
E_\text{Schott} &= E_\text{Schott}^\text{inst} + E_\text{Schott}^\text{tail} + E_\text{Schott}^\text{SO} + E_\text{Schott}^\text{SS}.
\end{align}
Note that when taking the time derivative of these Schott terms using the EOMs, the LO nonspinning part contributes to the LO SO and SS parts of the RR force.

\subsection{Solving for the eccentric-orbits RR force}
Using the fluxes and the Schott terms, the RR force can be calculated from the balance equations \eqref{balance}, which fix some of the unknowns in the ansatz for the Schott terms. The remaining unknowns can be determined by requiring that the RR force satisfies the conditions~\eqref{Fconds} in the circular-orbits limit.

\subsubsection{Leading order}
At leading order, calculating the RR force with the ansatz in Eqs.~\eqref{SchottInst} and expanding in $p_r$ gives
\begin{equation}
\mF_r = \frac{\nu^2}{5p_r r^3} \left(p^2-\frac{1}{r}\right) \left[p^2 \left(5 \alpha _1-5 \beta _2+16\right)-5 \frac{\beta _3}{r} \right]  + \Order(p_r).
\end{equation}
Requiring that the $1/p_r$ term is zero, leads to the solution
\begin{equation}
\beta_3 = 0, \qquad \alpha_1 = \frac{1}{5} (5 \beta_2-16).
\end{equation}
Expanding $\mF_r \pphi / (\mF_\phi p_r) - 1$ in $p_r$ yields
\begin{equation}
\frac{\mF_r \pphi}{\mF_\phi p_r} - 1 = \frac{9 \left(5 \beta _1-8\right) p^2-\left(45 \beta _1+30 \beta _2+88\right) / r}{ 15 \beta _2 p^2+3\left(32-5 \beta _2\right) / r} + \Order(p_r^2).
\end{equation}
Requiring that the first term in that series expansion is zero gives the solution
\begin{equation}
\beta_1 = \frac{8}{5}, \quad \beta_2 = -\frac{16}{3}, \quad \alpha_1 = -\frac{128}{15}.
\end{equation}
With that solution, we obtain the LO RR force
\begin{align}
\mF_\phi^\text{LO} &= \frac{8\nu^2}{15r^3} \pphi \left(10 p^2-39 p_r^2-\frac{22}{r} \right), \nonumber\\
\mF_r^\text{LO} &= -\frac{16\nu^2}{15r^3} p_r \left(-5 p^2+12 p_r^2+\frac{11}{r} \right).
\end{align}
This force satisfies the conditions in Eq.~\eqref{Fconds} for circular orbits since
\begin{align}
\mF_\phi^\text{LO} &= - \Phi_J^\text{LO} - \frac{128}{15r^3} \pphi \nu ^2 \left(2 p_r^2-\dot{p}_r r\right), \nonumber\\
\frac{\mF_r^\text{LO} \pphi}{\mF_\phi^\text{LO} p_r} &= 1 + \frac{15 p_r^2}{10 p^2-39 p_r^2-22 /r}\,.
\end{align}

\subsubsection{1PN}
Following the same steps as above, we obtain the following solution for the unknowns at 1PN:
\begin{align}
\beta _8&=\frac{1}{15} \left(371-15 \alpha _3+268 \nu\right), \nonumber\\
\alpha _2&=\frac{1}{35} \left(35 \beta _5+108 \nu -22\right), \nonumber\\
\beta _6&=\frac{2}{315} \left(105 \alpha _3-1760 \nu +1743\right), \nonumber\\
\beta _9&=-\frac{32}{105} (4 \nu -1), \nonumber\\
\beta _7&=\alpha _3+\frac{2}{35} (4 \nu +93),\nonumber\\
\alpha _4&=\frac{893}{21}-\alpha _3+\frac{2924 \nu }{105},
\end{align}
with 3 arbitrary coefficients out of 9 coefficients at that order. To simplify the resulting expressions for the RR force, we choose to set all arbitrary coefficients to zero, i.e. $\alpha_3 = \beta_4 = \beta_5 = 0$, which yields the following 1PN contribution to the RR force:
\begin{align}
\mF_\phi^\text{1PN} &= \frac{\nu^2  \pphi}{105 c^2r^3} \bigg[
18 (4 \nu +93) p_r^2 p^2-6 (4 \nu +93) p^4 \nonumber\\ 
&\quad +180 (7 \nu -5) p_r^4 +(9780 \nu +19198) \frac{p_r^2}{r} \nonumber\\
&\quad -(484 \nu + 3833) \frac{p^2}{r}+ \frac{1684 \nu +6213}{r^2}
\bigg] , \nonumber\\
\mF_r^\text{1PN} &= \frac{\nu^2 p_r}{105 c^2 r^3} \bigg[
180 (7 \nu -5) p_r^4 -6 (4 \nu +93) p^4 \nonumber\\
&\quad +4 (691 \nu +3958) \frac{p_r^2}{r} -198 (6 \nu -13) p_r^2 p^2 \nonumber\\
&\quad -(484 \nu + 3833) \frac{p^2}{r} + \frac{1684 \nu +6213}{r^2}
\bigg].
\end{align}

\subsubsection{LO tail}
Solving for the unknowns at the LO tail contribution leads to the solution
\begin{gather}
\lambda_2 = -\frac{334}{15}, \quad \lambda_3 = \frac{128}{5}, \quad 
\lambda_4 = -\frac{334}{15} + \lambda_1, \nonumber\\
\lambda_5 = -\frac{334}{15}, \quad \lambda_6 = \frac{128}{5},
\end{gather}
with either  $\lambda_1$ or $\lambda_4$ arbitrary.
Choosing $\lambda_1 = 0$, the tail contribution to the RR forces becomes
\begin{align}
\mF_\phi^\text{tail} &= \frac{\pi  \nu ^2}{c^3r^2}  \bigg(\frac{334 p^4}{15 r}-\frac{718 p^2}{15 r^2}-\frac{334 p^2 p_r^2}{5 r}-\frac{308 p_r^2}{15 r^2} \nonumber\\
&\quad +\frac{49 p_r^6}{225}\bigg), \nonumber\\
\mF_r^\text{tail} &= \frac{\pi \nu^2 \pphi p_r}{c^3r^4} \! \left(\!\frac{334 p^2}{15 r}-\frac{p_r^4}{18}-\frac{7034 p_r^2}{45 r}-\frac{718}{15 r^2}\!\right).
\end{align}

\subsubsection{LO spin-orbit}
At LO SO, we obtain the solution
\begin{align}
\sigma _5&=\frac{1}{15} \left(-36 \delta -32 \nu +15 \sigma _{11}+36\right), \nonumber\\
\sigma _6&=\frac{1}{15} \left(-68 \delta -40 \nu +15 \sigma _{12}+68\right), \nonumber\\
\sigma _2&=\frac{1}{15} \left(36 \delta -32 \nu +15 \sigma _8+36\right), \nonumber\\
\sigma _3&=\frac{1}{15} \left(68 \delta -40 \nu +15 \sigma _9+68\right), \nonumber\\
\sigma _{12}&=\frac{8}{15} (4 \delta +3 \nu -4), \nonumber\\
\sigma _9&=-\frac{8}{15} (4 \delta -3 \nu +4),\nonumber\\
\sigma _4&=\frac{1}{15} \left(-6 \delta +44 \nu +15 \sigma _{10}+6\right), \nonumber\\
\sigma _1&=\frac{1}{15} \left(6 \delta +44 \nu +15 \sigma _7+6\right), \nonumber\\
\sigma _{11}&=-\frac{2}{15} (9 \delta +22 \nu -9), \nonumber\\
\sigma _8&=\frac{2}{15} (9 \delta -22 \nu +9)
\end{align}
where either $\sigma_1$ or $\sigma_7$ is arbitrary, and either $\sigma_4$ or $\sigma_{10}$ is arbitrary. 
Choosing $\sigma_7 = \sigma_{10} = 0$, we obtain
\begin{align}
\mF_\phi^\text{SO} &= \frac{2\nu^2}{15c^3r^3} \chi_1\bigg[
p^4 (22 \nu-9 \delta -9) + 5 (3 + 3 \delta +16 \nu) p_r^4 \nonumber\\
&\quad 
+ (66 \nu -23 \delta -23) \frac{ p_r^2}{r} 
+(179 + 179 \delta -146 \nu )\frac{  p^2}{r}\nonumber\\
&\quad
+ 6 (9+9 \delta -22 \nu ) p^2 p_r^2 
\bigg] + 1\leftrightarrow 2. \\
\mF_r^\text{SO} &= \frac{2\nu^2p_r\pphi}{15 c^3 r^5} \chi_1 \bigg[
\frac{179 \delta -146 \nu +179}{r} + (22 \nu-9 \delta -9) p^2\nonumber\\
&\quad  -5 (3+3 \delta +16 \nu) p_r^2 \bigg] + 1\leftrightarrow 2.
\end{align}

\subsubsection{2PN no spin}
At 2PN, we obtain
\begin{align}
\alpha _6&=\beta _{13}-\frac{463 \nu ^2}{63}-\frac{1909 \nu }{315}+\frac{922}{315},\nonumber\\
\beta _{15}&=-\frac{2 \beta _{18}}{3}+\frac{1060 \nu ^2}{189}-\frac{12116 \nu }{189}-\frac{347236}{2835}, \nonumber\\
\beta _{16}&=-\beta _{17}-\beta _{18}-\frac{5276 \nu ^2}{315}-\frac{109609 \nu }{630}+\frac{9175}{378},\nonumber\\
\beta _{14}&=\alpha _7-\frac{2 \beta _{17}}{3}-\frac{2 \beta _{18}}{3}+\frac{244 \nu ^2}{45}-\frac{12466 \nu }{135}-\frac{28204}{2835}, \nonumber\\
\beta _{19}&=-\frac{640 \nu ^2}{189}+\frac{416 \nu }{35}-\frac{2608}{945}, \nonumber\\
\alpha _8&=-\beta _{17}-\beta _{18}-\frac{5018 \nu ^2}{315}-\frac{105491 \nu }{630}+\frac{52223}{1890},\nonumber\\
\alpha _{10}&=\beta _{18}-\frac{92 \nu ^2}{9}-\frac{309 \nu }{5}-\frac{10019}{315},\nonumber\\
\alpha _9&=\beta _{17}-\frac{78 \nu ^2}{7}+\frac{1201 \nu }{90}+\frac{5711}{126},
\end{align}
with 8 arbitrary coefficients out of 16. Choosing $\alpha _5 = \alpha _7= \beta _{10}= \beta _{11}= \beta _{12}=\beta _{13}=\beta _{17}= \beta _{18}= 0$ yields
\begin{widetext}
\begin{align}
\mF_\phi^\text{2PN} &= \frac{\nu^2 \pphi}{c^4r^3}\bigg\lbrace\!\!
\left(\frac{5276 \nu ^2}{315}+\frac{109609 \nu }{630}-\frac{9175}{378}\right)\! p^6
+\left(\frac{1519}{54}-\frac{9964 \nu ^2}{315}-\frac{100847 \nu }{630}\right)\! \frac{p^4}{r}
+\left(\frac{3512 \nu ^2}{315}-\frac{4234 \nu }{45}+\frac{3355}{21}\right)\! \frac{ p^2}{r^2}  \nonumber\\
&\quad 
+\left(\frac{9175}{126}-\frac{5276 \nu ^2}{105}-\frac{109609 \nu }{210}\right) p^4p_r^2 
+ \left(\frac{104296}{945}-\frac{15121 \nu ^2}{105}-\frac{254732 \nu }{315}\right)\frac{p_r^2 p^2}{r}
+\left(\frac{52}{3}-\frac{152 \nu ^2}{9}-\frac{310 \nu }{9}\right) p_r^6 \nonumber\\
&\quad
+\left(-\frac{185 \nu ^2}{63}+\frac{2171 \nu }{63}-\frac{269}{63}\right) p_r^4 p^2
+\left(\frac{4112}{35}-\frac{278 \nu ^2}{45}+\frac{344 \nu }{35}\right)\frac{p_r^4}{r} 
+\left(-\frac{1604 \nu ^2}{315}-\frac{2054 \nu }{7}-\frac{8803}{21}\right)\frac{p_r^2}{r^2}  \nonumber\\
&\quad  + \left(\frac{8 \nu ^2}{15}+\frac{9728 \nu }{315}-\frac{190244}{2835}\right)\frac{1}{r^3}
\bigg\rbrace, \\
\mF_r^\text{2PN} &= \frac{\nu^2 p_r}{c^4 r^3}\bigg\lbrace\!\!
\left(\frac{5276 \nu ^2}{315}+\frac{109609 \nu }{630}-\frac{9175}{378}\right)\! p^6
+\left(\frac{1519}{54}-\frac{9964 \nu ^2}{315}-\frac{100847 \nu }{630}\right)\! \frac{ p^4}{r}
+\left(\frac{3512 \nu ^2}{315}-\frac{4234 \nu }{45}+\frac{3355}{21}\right)\! \frac{ p^2}{r^2}\nonumber\\
&\quad 
+\left(\frac{9713}{126}-\frac{2129 \nu ^2}{45}-\frac{350537 \nu }{630}\right) p^4 p_r^2 
+\left(\frac{26459}{945}-\frac{1745 \nu ^2}{63}-\frac{449227 \nu }{315}\right)\frac{p_r^2 p^2}{r} 
+\left(\frac{52}{3}-\frac{152 \nu ^2}{9}-\frac{310 \nu }{9}\right) p_r^6 \nonumber\\
&\quad 
+\left(\frac{293 \nu ^2}{21}+\frac{1447 \nu }{21}-\frac{1361}{63}\right) p_r^4 p^2
+ \left(\frac{41612}{315}-\frac{1894 \nu ^2}{45}+\frac{28648 \nu }{315}\right)\frac{p_r^4}{r} 
+ \left(\frac{34528 \nu ^2}{945}-\frac{481624 \nu }{945}-\frac{18761}{45}\right)\frac{p_r^2}{r^2}  \nonumber\\
&\quad
+ \left(\frac{8 \nu ^2}{15} +\frac{9728 \nu }{315}-\frac{190244}{2835}\right)\frac{1}{r^3}
\bigg\rbrace.
\end{align}
\end{widetext}

\subsubsection{LO spin-spin}
At LO SS, we obtain the unique solution
\begin{align}
&\zeta _1=\frac{1}{30} \left[\delta  (73-128 \nu )+128 \nu ^2-274 \nu +73\right], \nonumber\\
&\zeta _2=-\frac{42}{5} (\delta -2 \nu +1), \quad \zeta _3=\frac{2}{15} \nu  (64 \nu -261), \nonumber\\
&\zeta _4=\frac{1}{30} \left[\delta  (128 \nu -73)+128 \nu ^2-274 \nu +73\right], \nonumber\\
&\zeta _5=\frac{42}{5} (\delta +2 \nu -1), \nonumber\\
&\zeta _6=-\frac{74}{15} \left(-2 \delta  \nu +\delta +2 \nu ^2-4 \nu +1\right), \nonumber\\
&\zeta _7=\frac{1}{10} \left[\delta  (43-80 \nu )+80 \nu ^2-166 \nu +43\right], \nonumber\\
&\zeta _8=0, \qquad \zeta _9=2 (\delta -2 \nu +1),\nonumber\\
&\zeta _{10}=-6 (\delta -2 \nu +1), \qquad \zeta _{11}=0, \nonumber\\
&\zeta _{12}=\frac{8}{15} (15-37 \nu ) \nu , \qquad \zeta _{13}=\frac{2}{5} \nu  (40 \nu -63), \nonumber\\
&\zeta _{14}=0, \qquad \zeta _{15}=-\frac{74}{15} \left[\delta  (2 \nu -1)+2 \nu ^2-4 \nu +1\right], \nonumber\\
&\zeta _{16}=\frac{1}{10} \left[\delta  (80 \nu -43)+80 \nu ^2-166 \nu +43\right], \nonumber\\
&\zeta _{17}=0, \qquad \zeta _{18}=-2 (\delta +2 \nu -1), \nonumber\\
&\zeta _{19}=6 (\delta +2 \nu -1), \qquad \zeta _{20}=0.
\end{align}
With that solution, we get
\begin{align}
\mF_\phi^\text{SS} &= \frac{\nu^2 \pphi}{30c^4r^5} \bigg\lbrace
\chi_1^2 \bigg[ 24X_1^2\CESA \left(15 p^2-90 p_r^2-\frac{49}{r}\right) \nonumber\\
&\quad\quad + \left(361 \delta-632 \nu\delta +632 \nu ^2-1354 \nu +361\right) p_r^2 \nonumber\\
&\quad\quad +\left(400 \nu\delta   -209 \delta -400 \nu ^2+818 \nu -209\right) p^2 \nonumber\\
&\quad\quad +\left(355 \delta-704 \nu\delta   +704 \nu ^2-1414 \nu +355\right) \frac{1}{r} \bigg]\nonumber\\
&\quad+ \nu\chi_1\chi_2 \bigg[ (378-400 \nu ) p^2+(632 \nu -2250) p_r^2 \nonumber\\
&\quad\quad +\frac{704 \nu -1182}{r}\bigg]  + 1\leftrightarrow 2
\bigg\rbrace, \\
\mF_r^\text{SS} &= \frac{\nu^2 p_r}{30c^4 r^5} \bigg\lbrace \chi_1^2 \bigg[
24X_1^2\CESA \left(15 p^2-55 p_r^2-\frac{49}{r}\right) \nonumber\\
&\quad\quad + \left(301 \delta -512 \delta  \nu +512 \nu ^2-1114 \nu +301\right) p_r^2  \nonumber\\
&\quad\quad + \left(400 \nu\delta -209 \delta -400 \nu ^2+818 \nu -209\right) p^2 \nonumber\\
&\quad\quad + \left(355 \delta -704 \delta  \nu+704 \nu ^2-1414 \nu +355\right) \frac{1}{r} \bigg] \nonumber\\
&\quad +\nu\chi_1\chi_2 \bigg[
(378-400 \nu ) p^2+(512 \nu -1410) p_r^2 \nonumber\\
&\quad\quad +\frac{704 \nu -1182}{r} \bigg]  + 1\leftrightarrow 2
\bigg\rbrace.
\end{align}

\subsection{Factorizing the RR force into circular and noncircular parts}
The total RR force is the sum of the contributions calculated in the previous section, i.e.,
\begin{align}
\mF_\phi &= \mF_\phi^\text{LO} + \mF_\phi^\text{1PN} + \mF_\phi^\text{2PN} + \mF_\phi^\text{tail} + \mF_\phi^\text{SO} + \mF_\phi^\text{SS}, \nonumber\\
\mF_r &= \mF_r^\text{LO} + \mF_r^\text{1PN} + \mF_r^\text{2PN} + \mF_r^\text{tail} + \mF_r^\text{SO} + \mF_r^\text{SS}.
\end{align}
We have checked that our gauge-dependent RR force agrees with that in Refs.~\cite{Bini:2012ji,Zeng:2007bq,Wang:2007ntb} by using the balance equations.
Denoting the RR force from those references by $(\bar{\mathcal{F}}_r, \bar{\mathcal{F}}_\phi)$ with corresponding Schott terms $(\bar{E}_\text{Schott}, \bar{J}_\text{Schott})$, Eqs.~\eqref{baleqns} lead to
\begin{align}
\dot{\bar{r}} \bar{\mathcal{F}}_r + \dot{\bar{\phi}} \bar{\mathcal{F}}_\phi + \dot{\bar{E}}_\text{Schott} &= \dot{r} \mathcal{F}_r + \dot{\phi} \mathcal{F}_\phi + \dot{E}_\text{Schott}, \nonumber\\
\bar{\mathcal{F}}_\phi + \dot{\bar{J}}_\text{Schott} &= \mathcal{F}_\phi + \dot{J}_\text{Schott}.
\end{align}
Then, by writing an ansatz for $(\bar{E}_\text{Schott}, \bar{J}_\text{Schott})$ with unknown coefficients, we checked that a solution exists, implying that $(\mathcal{F}_r,\mathcal{F}_\phi)$ and $(\bar{\mathcal{F}}_r,\bar{\mathcal{F}}_\phi)$ are related via a coordinate transformation.

To implement our results in the \texttt{SEOBNRv4HM} model, we factorize the RR force into a quasi-circular part times eccentric corrections as in Eqs.~\eqref{Ffact} and~\eqref{RRFcirc}, which read
\begin{align}
\label{RRFqc}
\mF_\phi &= \mF_\phi^\text{qc} \mF_\phi^\text{ecc}, \qquad \mF_r = \mF_r^\text{qc} \mF_r^\text{ecc}, \nonumber\\
\mF_\phi^\text{qc} &= - \frac{\Phi_E^\text{qc}}{\Omega}, \qquad
\mF_r^\text{qc} = -  \frac{\Phi_E^\text{qc} p_r}{\Omega \pphi},
\end{align}
and for the quasi-circular part we use the unexpanded force used in \texttt{SEOBNRv4HM}, in which the energy flux has the following PN expansion in terms of the orbital velocity $v_\Omega \equiv \Omega^{1/3}$:
\begin{align}
\frac{\Phi_E^\text{qc}}{\nu^2} &= \frac{32}{5} v_\Omega^{10} - \frac{2}{105 c^2} (980 \nu +1247) v_{\Omega }^{12} + \frac{128 \pi }{5c^3}v_{\Omega }^{13} \nonumber\\
&\quad + \frac{4v_{\Omega }^{13}}{5c^3} \left[\chi _1 (12 \nu-11 \delta -11)  + \chi _2 (12 \nu + 11 \delta -11)\right] \nonumber\\
&\quad
+ \frac{2}{2835 c^4} v_{\Omega }^{14}\left(32760 \nu ^2+166878 \nu -44711\right) \nonumber\\
&\quad
+ \frac{2}{5c^4} v_{\Omega }^{14} \big[(32 \CESA+1) \chi _1^2 X_1^2+62 \nu  \chi _1\chi _2 \nonumber\\
&\quad\qquad + (32 \CESB+1) \chi _2^2 X_2^2\big].
\end{align}
This leads to the eccentric part
\begin{align}
\mF_\phi^\text{ecc} &= \frac{29 p_r^2 r-10 \dot{p}_r r^2+12}{12 \left(\dot{p}_r r^2+1\right)^{2/3}} + \frac{[\dots]}{c^2 r \left(\dot{p}_r r^2+1\right)^{5/3}}  + \dots \nonumber\\
\mF_r^\text{ecc} &= \frac{7 p_r^2 r-5 \dot{p}_r r^2+6}{6 \left(\dot{p}_r r^2+1\right)^{2/3}} + \frac{[\dots]}{c^2 r \left(\dot{p}_r r^2+1\right)^{5/3}}  + \dots
\end{align}
The full 2PN expressions are provided in the Supplemental Material~\cite{ancmaterial}.

In these eccentric corrections to the RR force, we used $\dot{p}_r$ instead of $\pphi$ because it improves the agreement of our model with \texttt{SEOBNRv4HM} in the quasi-circular orbit limit, in which $p_r = 0 = \dot{p}_r$ leading to $\mF_{r,\phi}^\text{ecc} = 1$. However, having $\dot{p}_r$ on the right-hand side of the EOM for $p_r$ would complicate solving the system of differential equations~\eqref{EOMs}. Therefore,  when evolving the EOMs, we simply replace $\dot{p}_r$ in the RR force with the derivative of the Hamiltonian with respect to $r$ calculated numerically, i.e. $\dot{p}_r \to -\partial \hat{H}_\text{EOB}/\partial r$.

SEOBNR waveform models use $p_{r_*}$ (the conjugate momentum to the tortoise radial coordinate $r_*$) instead of $p_r$ since it improves stability of the EOMs near the EOB event horizon~\cite{Damour:2007xr,Pan:2009wj}. The two momenta are related by Eq.~\eqref{prstar}. In Appendix~\ref{app:tort}, we also obtain Eq.~\eqref{prdotstar} for the transformation between $\dot{p}_r$ and $\dot{p}_{r_*}$.

\section{Initial conditions}
\label{sec:initCond}
Having determined the RR force that enters the EOMs, we need to specify the initial conditions to be used in evolving the system of equations. In this section, we first review how the initial conditions are implemented in \texttt{SEOBNRv4HM} for quasi-circular orbits~\cite{Buonanno:2000ef,Buonanno:2005xu}, and then discuss a simple extension for eccentric orbits.

\subsection{Initial conditions for quasi-circular orbits}
\label{initialcirc}

Let us recapitulate the initial conditions for quasi-circular/spherical orbits in the \texttt{SEOBNRv4HM} model as derived in Refs.~\cite{Buonanno:2000ef,Buonanno:2005xu}.
We start by specifying an initial orbital frequency $\Omega_0$, with initial orbital phase $\phi_0 = 0$, and solve
\begin{equation}
\label{circICrL}
\left[\frac{\partial H}{\partial r}\right]_0 = 0, \qquad \left[\frac{\partial H}{\partial \pphi}\right]_0 = \Omega_0
\end{equation}
for the initial values of $r$ and $\pphi$, while neglecting RR, $p_r \approx 0$. The initial condition for $p_r$ is then obtained by solving
\begin{equation}
[\dot{r}]_0 = \left[\frac{\partial H}{\partial p_r}\right]_0
\end{equation}
for $p_r$, after calculating $[\dot{r}]_0$ using the result from adiabatic evolution~\cite{Buonanno:2000ef}
\begin{equation}
[\dot{r}]_0 = \left[\frac{d\pphi/dt}{d\pphi/dr} \right]_0 = \left[\frac{\dot{E}}{dE/dr}\right]_0,
\end{equation}
where $\dot{E}$ is the circular-orbits energy flux, and the derivative $dE/dr = dH/dr$ can be determined using the following equations for circular orbits:
\begin{gather}
dH = \frac{\partial H}{\partial r} dr + \frac{\partial H}{\partial p_r} dp_r + \frac{\partial H}{\partial \pphi} d\pphi, \\
p_r = 0, \quad  dp_r = 0, \quad d\left(\frac{\partial H}{\partial r}\right) = 0.
\end{gather}
This leads to
\begin{gather}
\frac{d}{dr}\frac{\partial H}{\partial r} = 0 = \frac{\partial^2 H}{\partial r^2} + \frac{\partial^2 H}{\partial r \partial \pphi} \frac{d\pphi}{dr},
\end{gather}
which can be solved for $d\pphi/dr$ to obtain
\begin{equation}
\frac{d\pphi}{dr} = - \frac{\partial^2 H / \partial r^2}{\partial^2 H / \partial r\partial \pphi}\,.
\end{equation}
Plugging that solution into $dH/dr = (\partial H / \partial \pphi) d\pphi/dr$ yields the result in Eq.~(4.14) of Ref.~\cite{Buonanno:2005xu}, which reads
\begin{equation}
\frac{dH}{dr} = - \frac{(\partial H /\partial \pphi) (\partial^2 H / \partial r^2)}{\partial^2 H / \partial r\partial \pphi},
\end{equation}
and hence
\begin{equation}
\label{rdotcirc}
\left[\frac{\partial H}{\partial p_r}\right]_0 = [\dot{r}]_0 = \left[- \frac{dE}{dt} \frac{\partial^2 H / \partial r\partial \pphi}{(\partial H /\partial \pphi) (\partial^2 H / \partial r^2)} \right]_0.
\end{equation}

The complete procedure to obtain the initial conditions for the orbital phase-space is now as follows.
Given $\Omega_0$, masses, and spins, we numerically solve the relations in Eq.~\eqref{circICrL} for the initial values $r_0$ and ${\pphi}_0$, choosing $\phi_0=0$ and assuming $p_r \approx 0$.
Using these values, we numerically solve Eq.~\eqref{rdotcirc} for the initial value $p_{r0}$.

\subsection{Initial conditions for eccentric orbits}
Since eccentricity is a gauge-dependent concept, we do not need to calculate accurate initial conditions for eccentric orbits in a specific gauge. Instead, we can choose a measure for eccentricity that can be adjusted to be as convenient as possible for numerical implementation. The only strict requirement is that for zero eccentricity $e=0$ one recovers the quasi-circular case. Hence, we can start with very accurate initial conditions for quasi-circular orbits and perturb them for eccentric orbits.

We choose to specify an initial orbital frequency $\Omega_0$ and an initial eccentricity $e_0$ using the Keplerian parametrization $1/r = u_p (1 + e \cos\chi)$. We also assume that the orbit starts with $\phi_0 = 0$ at \emph{periastron} $(\chi = 0)$, where $p_r = 0$ in absence of RR, which simplifies calculating the initial conditions for $r$ and $\pphi$.
An advantage of starting at periastron instead of apastron is that the specified initial frequency is then the maximum orbital frequency (over the first orbit), and can be used to estimate the frequency at which the binary enters a GW detector's frequency band.

To obtain $r_0$ and ${\pphi}_0$, we solve Eq.~\eqref{circICrL} with a nonzero $\dot{p}_r$, i.e.,
\begin{equation}
\label{eccICrL}
\left[\frac{\partial H}{\partial r}\right]_0 = - \left[\dot p_r(\pphi,e)\right]_0, \qquad \left[\frac{\partial H}{\partial \pphi}\right]_0 = \Omega_0,
\end{equation}
with $p_r \approx 0$, and $[\dot p_r]_0$ given as a 2PN expansion in terms of $\pphi$ and $e$. For quasi-circular orbits, these equations reduce exactly to Eqs.~\eqref{circICrL} since $\dot p_r \propto e$. 

To obtain the PN expansion for $\dot p_r$ at periastron, we first invert the Hamiltonian at the turning points $r_\pm = 1 / (u_p(1 \pm e))$ with $p_r = 0$ and solve for the energy and angular momentum as functions of $e$ and $u_p$, which are given by Eqs.~\eqref{ELeup}. Then, we invert $\pphi(e, u_p)$ to obtain Eq.~\eqref{upLe} for $u_p(\pphi,e)$ and insert it into the PN expansion for $\dot{p}_r = -\partial H / \partial r$ at periastron ($r = 1 / [u_p (1+e)]$). This yields
\begin{align}
  \label{eq:prdot0}
[\dot p_r]_0 &=\frac{e_0 (e_0+1)^2}{p_{\phi 0}^4} 
+ \frac{e_0 (e_0+1)^2 }{2 c^2 p_{\phi 0}^6} \! \left[e_0^2 (5 -\nu)-4 e_0+\nu +7\right]\nonumber\\
&\quad + \frac{e_0 (e_0+1)^2 \left(3 e_0^2-2 e_0+3\right) }{2 c^3 p_{\phi 0}^7} \Big[(\nu-2 \delta  -2)\chi_1 \nonumber\\
&\quad\quad  +(\nu + 2 \delta -2)\chi_2 \Big]  +\frac{e_0 (e_0+1)^2}{8 c^4 p_{\phi 0}^8} \Big[
3 \nu ^2-\nu + 95 \nonumber\\
&\quad\quad -6 e_0^2 \left(\nu ^2+5 \nu -39\right)  +8 e_0 (\nu -25) \nonumber\\
&\quad\quad + e_0^4 \left(3 \nu ^2-17 \nu +55\right)+8 e_0^3 (\nu -7)\Big] \nonumber\\
&\quad + \frac{e_0 (e_0+1)^2}{2 c^4p_{\phi 0}^8} \Big\lbrace\chi _1^2 \Big[ \CESA \left(3 e_0^2-2 e_0+3\right) X_1^2 \nonumber\\
&\quad\quad +\left(5 e_0^2-6 e_0-3\right) X_1^4\Big] + \nu  \chi _1 \chi _2 \Big[e_0^2 (5 \nu +3) \nonumber\\
&\quad\quad -2 e_0 (3 \nu +1)-3 \nu +3\Big] + 1 \leftrightarrow 2\Big\rbrace.
\end{align}

The initial condition can now be obtained in analogy to the quasi-circular case:
given $\Omega_0$, $e_0$, masses, and spins, we obtain $r_0$ and ${\pphi}_0$  from Eqs.~\eqref{eccICrL} and \eqref{eq:prdot0} (assuming $p_r \approx 0$), $p_{r0}$ then follows from Eq.~\eqref{rdotcirc} as before, and $\phi_0 = 0$ by convention.
We can keep using the circular-orbits energy flux in Eq.~\eqref{rdotcirc}, instead of replacing $\dot{E}$ with $\Omega \mF_\phi$ for eccentric orbits, because the difference on the orbital dynamics is negligible since it involves $p_r$ (which at periastron is numerically much smaller than $r$ or $\pphi$).

To assess the accuracy of the initial conditions for eccentric orbits, we compare the specified value for the eccentricity in the Keplerian parametrization with the value calculated from the orbital frequency (or the frequency of the $(2,2)$ mode) at periastron $\Omega_p$ and apastron $\Omega_a$, which is given by \cite{Mora:2002gf}
\begin{equation}
\label{eomega}
e_\Omega = \frac{\sqrt{\Omega_p} - \sqrt{\Omega_a}}{\sqrt{\Omega_p} + \sqrt{\Omega_a}}\,,
\end{equation}
and to calculate it, we follow the steps explained after Eq. (2.8) of Ref.~\cite{Ramos-Buades:2019uvh}. Since we evaluate $e_\Omega$ by evolving the binary over one orbit (including RR), it only holds approximately that $\Omega_p \approx \Omega_a$ in the quasi-circular case. That is, $e_\Omega$ does not vanish exactly for quasi-circular orbits, in contrast to our $e_0$. Table~\ref{tab:eomega} shows the value of the eccentricity $e_\Omega$ calculated from the orbital frequency compared to the specified eccentricity $e_0$, and we see good agreement between the two measures of eccentricity.

\begin{table}[h]
\caption{Nonspinning initial conditions given the parameters ($e_0,\Omega_0,\nu$), and the eccentricity $e_\Omega$ measured from the orbital frequency using Eq.~\eqref{eomega}. The initial frequency $\Omega_0$ was chosen to give $\sim 30$ GW cycles between $r_0$ and $r = 5$.}
\begin{ruledtabular}
\begin{tabular}{ccc|cccc}
$e_0$ & $\Omega_0$ & $\nu$ & $r_0$ & ${\pphi}_0$ & ${p_r}_0$ & $e_\Omega$ \\
\hline
0.01 & 0.03 & 0.25 & 10.4 & 3.8 & -0.0012 & 0.0087 \\
0.2 & 0.045 & 0.25 & 8.1 & 3.7 & -0.0024 & 0.19 \\
0.7 & 0.065 & 0.25 & 6.7 & 4.0 & -0.0033 & 0.69 \\
0.2 & 0.058 & 0.1 & 6.8 &  3.6 & -0.0011 & 0.20 \\
\end{tabular}
\end{ruledtabular}
\label{tab:eomega}
\end{table}

\section{Gravitational waveform modes}
\label{sec:modes}
In this section, we obtain the 2PN waveform modes to 2PN order including LO tail effects, and SO and SS couplings for aligned spins. The instantaneous nonspinning part of the modes was derived for eccentric orbits in Refs.~\cite{Gopakumar:2001dy,Mishra:2015bqa} in harmonic coordinates and we convert their results to EOB coordinates.
For the LO tail part, we extend the results of Ref.~\cite{Hinderer:2017jcs} to $\Order(e^6)$ and to higher modes using the Keplerian parametrization, and then convert those results to an expansion in $p_r$ and $\dot{p}_r$.
The spin contributions to the modes were derived to 2PN order for circular orbits in Ref.~\cite{Buonanno:2012rv}, and here we derive them for eccentric orbits.

The GW spherical harmonic modes $h^{\ell m}$ are the expansion of the complex polarization waveform $h = h_+ - i h_\times$ into spin-weighted $s = -2$ spherical harmonics $Y^{\ell m}_{-2}(\Theta,\Phi)$ such that
\begin{equation}
h_+ - i h_\times = \sum_{\ell = 2}^{\infty} \sum_{m=-\ell}^{\ell} h^{\ell m} Y^{\ell m}_{-2}(\Theta,\Phi).
\end{equation}
The modes $h^{\ell m}$ can be calculated directly from the radiative multipole moments via~\cite{Kidder:2007rt,Blanchet:2008je,Faye:2012we}
\begin{equation}
\label{hlm}
h^{\ell m} = \frac{1}{\sqrt{2} D_L c^{\ell + 2}} \left[U^{\ell m} - \frac{i}{c} V^{\ell m}\right],
\end{equation}
where $D_L$ is the luminosity distance of the source, and the  radiative multipole moments are related to the symmetric trace-free (STF) moments $U_L$ and $V_L$ by
\begin{align}
U^{\ell m} &= \frac{16\pi}{(2\ell + 1)!!} \sqrt{\frac{(\ell + 1)(\ell + 2)}{2\ell (\ell - 1)}} \bar{\mathcal{Y}}_L^{\ell m} U_L, \nonumber\\
V^{\ell m} &= \frac{-32\pi}{(2\ell + 1)!!} \sqrt{\frac{\ell(\ell + 2)}{2(\ell + 1) (\ell - 1)}} \bar{\mathcal{Y}}_L^{\ell m} V_L, \label{Vlm}
\end{align}
where $\bar{\mathcal{Y}}_L^{\ell m}$ is the complex conjugate of the STF tensors relating the unit vectors $N_{\langle L\rangle}$ (which point from the source to the detector) to the spherical harmonics basis $Y^{\ell m}(\Theta,\Phi)$ such that
\begin{align}
Y^{\ell m}(\Theta,\Phi) &= \mathcal{Y}_L^{\ell m} N_{\langle L \rangle} (\Theta,\Phi), \\
\label{NYlm}
N_{\langle L\rangle}(\Theta,\Phi) &= \frac{4\pi \ell!}{(2\ell + 1)!!} \sum_{m=-\ell}^{\ell} \bar{\mathcal{Y}}_L^{\ell m} Y^{\ell m}(\Theta,\Phi), \\
\bar{\mathcal{Y}}_L^{\ell m} &= \frac{(2\ell + 1)!!}{4\pi \ell!} \int d\Omega\, N_{\langle L\rangle} \bar{Y}^{\ell m}, \\
\bar{\mathcal{Y}}_L^{\ell m} \mathcal{Y}_L^{\ell m'} &= \frac{(2\ell + 1)!!}{4\pi \ell!} \delta_{mm'},
\end{align}
and we can express the unit vector $\bm{N}$ in terms of the angles $\Theta$ and $\Phi$ as 
\begin{equation}
\bm{N} = \sin \Theta \cos \Phi \hat{\bm{e}}_x +  \sin \Theta \sin \Phi \hat{\bm{e}}_y +  \cos \Theta \hat{\bm{e}}_z.
\end{equation}
For planar binaries, nonspinning or with aligned spins, it was shown in Ref.~\cite{Faye:2012we} that the modes can be determined using the mass-type multipole moments for even $\ell + m$, or the current-type multipole moments for odd $\ell + m$, i.e.,
\begin{alignat}{3}
& h^{\ell m}   && = \frac{1}{\sqrt{2} D_L c^{\ell + 2}} U^{\ell m}, \qquad && \ell + m \text{ even} \nonumber\\
& h^{\ell m}   && = -\frac{i}{\sqrt{2} D_L c^{\ell + 3}} V^{\ell m}, \qquad && \ell + m \text{ odd.}
\end{alignat}
We define $H^{\ell m}$ such that
\begin{equation}
h^{\ell m} = -\frac{8 \mu}{c^4 D_L} \sqrt{\frac{\pi}{5}} \mathrm{e}^{-im\phi} H^{\ell m},
\end{equation}
which makes the LO part of $H^{22} = x$ for circular orbits.
Note that different conventions for the phase origin contribute a factor of $(-i)^m$ to the modes \cite{Arun:2008kb}. 

In this paper, we compute the modes to 2PN order beyond the leading order of the $(2,2)$ mode, which means we consider modes up to the $\ell = 6,~m = $ even modes.
To 2PN order, the instantaneous contributions to the radiative multipole moments coincide with the source multipole moments. Including the hereditary terms that contribute to 2PN, the radiative multipole moments are given by~\cite{Mishra:2015bqa,Blanchet:2008je,Kidder:2007rt}
\begin{align}
\label{radMoments}
U_{ij} &= I_{ij}^{(2)} + \frac{2 M}{c^3} \int_{0}^{\infty} d\tau \, I_{ij}^{(4)} (t - \tau) \ln\left(\frac{\tau}{b_1}\right) + \Order\left(\frac{1}{c^5}\right), \nonumber\\
U_{ijk} &= I_{ijk}^{(3)} + \frac{2 M}{c^3} \int_{0}^{\infty} d\tau \, I_{ijk}^{(5)} (t - \tau) \ln\left(\frac{\tau}{b_2}\right) + \Order\left(\frac{1}{c^5}\right), \nonumber\\
U_{L} &= I_{L}^{(\ell)}  + \Order\left(\frac{1}{c^3}\right), \nonumber\\
V_{ij} &= J_{ij}^{(2)} + \frac{2 M}{c^3} \int_{0}^{\infty} \! d\tau \, J_{ij}^{(4)} (t - \tau) \ln\left(\frac{\tau}{b_3}\right) + \Order\left(\frac{1}{c^5}\right), \nonumber\\
V_{L} &= J_{L}^{(\ell)} + \Order\left(\frac{1}{c^3}\right),
\end{align}
where the constants $b_i$ are gauge parameters that will be eliminated via a phase shift as was done in Ref.~\cite{Kidder:2007rt}. The source multipole moments for nonspinning binaries are given in, e.g., Refs.~\cite{Kidder:2007rt,Mishra:2015bqa}, while the spin contributions to the source moments are given in Refs.~\cite{Blanchet:2006gy,Buonanno:2012rv}.

\subsection{Instantaneous nonspinning contributions}
The instantaneous contributions to the modes for nonspinning binaries in eccentric orbits were derived in Ref.~\cite{Gopakumar:2001dy} to 2PN, and in Ref.~\cite{Mishra:2015bqa} to 3PN. The results of Ref.~\cite{Mishra:2015bqa} are in harmonic coordinates and in terms of the variables ($r,\phi,\dot{r},\dot{\phi}$). Hence, we can simply transform their results from harmonic to EOB coordinates using the transformations in Appendix~\ref{app:transform}. For the $(2,2)$ mode we obtain
\begin{widetext}
\begin{align}
\label{H22inst}
\hat{H}_\text{inst}^{22} &= \frac{1}{2} \left(\frac{1}{r} + p^2 - 2p_r^2\right) + i \frac{\pphi p_r}{r} 
+ \frac{1}{c^2} \Bigg\lbrace 
\left(\frac{\nu }{28}-\frac{5}{28}\right) p^4+\left(\frac{31 \nu }{28}-\frac{157}{84}\right)\frac{ p^2}{r}+\left(\frac{5}{14}-\frac{\nu }{14}\right) p_r^2 p^2+\left(\frac{13}{3}-\nu\right)\frac{p_r^2}{r} \nonumber\\
&\quad\qquad
+\left(\frac{\nu }{2}-2\right)\frac{1}{r^2} +i \frac{\pphi p_r}{r} \left[\left(\frac{\nu }{14}-\frac{5}{14}\right) p^2+ \left(\frac{2 \nu }{7}-\frac{185}{42}\right)\frac{1}{r}\right]
\Bigg\rbrace 
+ \frac{1}{c^4} \Bigg\lbrace 
\left(-\frac{17 \nu ^2}{336}-\frac{13 \nu }{336}+\frac{5}{48}\right) p^6 \nonumber\\
&\quad\qquad  
+\left(-\frac{671 \nu ^2}{1008}-\frac{1375 \nu }{1008}+\frac{481}{504}\right)\frac{ p^4}{r}+\left(\frac{127 \nu ^2}{54}-\frac{1355 \nu }{189}-\frac{5519}{3024}\right) \frac{ p^2}{r^2}+ \left(\frac{17 \nu ^2}{168}+\frac{13 \nu }{168}-\frac{5}{24}\right) p_r^2p^4 \nonumber\\
&\quad\qquad 
+\left(-\frac{67 \nu ^2}{126}+\frac{20 \nu }{9}-\frac{659}{504}\right) \frac{ p_r^2p^2}{r}+ \left(-\frac{464 \nu ^2}{189}+\frac{2249 \nu }{756}-\frac{811}{3024}\right)\frac{p_r^2}{r^2} 
+\left(\frac{17 \nu ^2}{18}-\frac{25 \nu }{36}-\frac{919}{1008}\right)\frac{ p_r^4}{r} \nonumber\\
&\quad\qquad
+ \left(\frac{205 \nu ^2}{252}-\frac{49 \nu }{36}+\frac{95}{63}\right)\frac{1}{r^3} 
+ i \frac{p_r \pphi}{r} \bigg[
\left(-\frac{17 \nu ^2}{168}-\frac{13 \nu }{168}+\frac{5}{24}\right) \frac{  p^4}{r} 
+\left(-\frac{4 \nu ^2}{21}+\frac{29 \nu }{28}+\frac{67}{56}\right)\frac{ p^2}{r^2}\nonumber\\
&\quad\qquad 
+\left(-\frac{523 \nu ^2}{378}+\frac{1226 \nu }{189}+\frac{193}{54}\right)\frac{1}{r^3} + \left(\frac{43 \nu ^2}{63}-\frac{125 \nu }{126}+\frac{787}{504}\right)\frac{ p_r^2}{r^2}
\bigg]
\Bigg\rbrace.
\end{align}
\end{widetext}
The expressions for the other modes that contribute to 2PN, i.e., up to $\ell=|m|=6$, are provided as a \textit{Mathematica} file in the Supplemental Material~\cite{ancmaterial}.
Note that the $(\ell,0)$ modes are zero for circular orbits but not for eccentric orbits. For example, the LO part of the $(2,0)$ mode is given by
\begin{equation}
\hat{H}_\text{inst}^{20} = \frac{1}{\sqrt{6}} \left(p^2 - \frac{1}{r}\right) + \Order\left(\frac{1}{c^2}\right),
\end{equation}
which is zero for circular orbits since $p^2 = 1/r + \dots$.

\subsection{Hereditary contributions}
\label{sec:modesTail}
The hereditary contributions to the modes can be calculated analytically in an eccentricity expansion, as was done in Ref.~\cite{Hinderer:2017jcs} for the $(2,2)$ mode to $\Order(e^2)$, and in Ref.~\cite{Boetzel:2019nfw} for all modes to 3PN order and to $\Order(e^6)$. 
The results of Ref.~\cite{Boetzel:2019nfw} use the quasi-Keplerian parametrization, while here we use the Keplerian parametrization following the method developed in Ref.~\cite{Hinderer:2017jcs}, which is based on results from Refs.~\cite{Drasco:2005is,Arun:2007rg,Kidder:2007rt}, to derive the leading order tail effects that contribute to the modes up to 2PN order and to $\Order(e^6)$. (See Ref.~\cite{Hinderer:2017jcs} for a discussion of the advantages of the Keplerian parametrization over the quasi-Keplerian parametrization.)
We finally convert the eccentricity-expanded tail contributions to an expansion in $p_r$ and $\dot{p}_r$.

\subsubsection{Modes with even $\ell + m$}
The LO mass-type multipole moments are given by~\cite{Arun:2007rg}
\begin{equation}
I^L = \mu s_\ell r^\ell n^{\langle L\rangle},
\end{equation}
where $s_\ell \equiv X_2^{\ell - 1} + (-1)^\ell X_1^{\ell - 1}$, and the unit vectors $n^{\langle L\rangle}$ are related to spherical harmonics via Eq.~\eqref{NYlm}, leading to
\begin{equation}
I^L = \sum_{m=-\ell}^{\ell} \mathcal{Y}_{\ell m}^L a_{\ell m} r^\ell \mathrm{e}^{-im\phi},
\end{equation}
with the coefficients (for equatorial orbits)
\begin{equation}
a_{\ell m} \equiv \frac{4\pi \mu s_\ell  \ell!}{(2\ell + 1)!!} \bar{Y}_{\ell m}\left(\frac{\pi}{2},0\right).
\end{equation}

Decomposing the phase $\phi$ into an oscillatory part and a linearly growing part $\phi = \phi_0 + \omega_\phi t + \Delta \phi_r$ allows expressing the oscillatory part $\Delta \phi_r$ as a Fourier series expansion. Hence, 
\begin{equation}
I^L = \sum_{m=-\ell}^{\ell} \mathcal{Y}_{\ell m}^L a_{\ell m} J_{\ell m} \mathrm{e}^{-im\psi_\phi},
\end{equation} 
with the functions $J_{\ell m}$ defined by
\begin{equation}
J_{\ell m} = r^\ell \mathrm{e}^{-im\phi_0} \mathrm{e}^{-im\Delta\phi_r} = \sum_{k=-\infty}^{\infty} J_{\ell mk} \mathrm{e}^{-ik\psi_r},
\end{equation}
where $\psi_r$ and $\psi_\phi$ are the radial and azimuthal angle variables associated with the frequencies $\omega_r=d\psi_r/dt$ and $\omega_\phi=d\psi_\phi/dt$. The coefficients $J_{\ell m k}$ are given by
\begin{align}
J_{\ell m k} &= \frac{1}{2\pi} \int_{0}^{2\pi} d\psi_r \, \mathrm{e}^{ik\psi_r} J_{\ell m} \nonumber\\
&= \frac{\omega_r}{2\pi} \int_{0}^{2\pi} \frac{d\chi}{\mathcal{P}} r^\ell \mathrm{e}^{-im\phi_0} \mathrm{e}^{-im\Delta\phi_r} \mathrm{e}^{ik\psi_r} \nonumber\\
&= 
\frac{\omega_r}{2\pi} 
u_p^{-\ell-3/2} \int_{0}^{2\pi} \frac{d\chi}{(1 + e\cos\chi)^{\ell +2}} \mathrm{e}^{-im\Delta\phi_r} \mathrm{e}^{ik\psi_r} ,
\end{align}
where, in the last line, we assume $\phi_0=0$. The function $\mathcal{P}$ denotes the conservative part of $\dot{\chi}$ and is related to the radial angle $\psi_r$ via $d\psi_r/d\chi = \omega_r / \mathcal{P}$, with $\mathcal{P} = (1+e\cos\chi)^2 u_p^{3/2}$ at LO (see Ref.~\cite{Hinderer:2017jcs} for more details).

Thus, the Newtonian mass multipole moments can be expressed as
\begin{equation}
I^L = \sum_{m=-\ell}^{\ell} \sum_{k=-\infty}^{\infty} \mathcal{Y}_{\ell m}^L a_{\ell m} J_{\ell m k} \mathrm{e}^{-i(k\psi_r + m \psi_\phi)},
\end{equation}
where the azimuthal angle is related to the radial angle by $\psi_\phi = \phi - \Delta\phi_r$ with $\Delta\phi_r = \chi - \psi_r$ at LO.
This allows us to write the LO tail contribution to the mass-type radiative moments as
\begin{align}
U_L^\text{tail} &= \frac{2M}{c^3} \int_{0}^{\infty} d\tau\, I_L^{(\ell+2)} (t - \tau) \ln\left(\frac{\tau}{b}\right), \nonumber\\
&= (-i)^{\ell + 2}\frac{2M}{c^3} \sum_{m=-\ell}^{\ell} \sum_{k=-\infty}^{\infty} 
\mathcal{Y}_L^{\ell m} a_{\ell m} J_{\ell m k} \Omega_{mk}^{\ell+2} \nonumber\\
&\quad\qquad \times \mathrm{e}^{-i(k\psi_r + m \psi_\phi)} \mathcal{I}(\Omega_{mk}),
\end{align}
where $\Omega_{mk} \equiv m \omega_\phi + k \omega_r$ and 
\begin{align}
\mathcal{I}(x) &\equiv \int_{0}^{\infty} d\tau \,  \mathrm{e}^{ix\tau} \ln\left(\frac{\tau}{b}\right) \nonumber\\
&= -\frac{1}{x} \left[\frac{\pi}{2} \text{sgn}(x) + i \ln(|x|b) + i \gamma_E\right].
\end{align} 
The exponential $\mathrm{e}^{-ik\psi_r}$ can be expressed in terms of $\chi$ and $e$ by integrating $d\psi_r = \omega_r d\chi /\mathcal{P}$, with $\omega_r = (u_p-e^2 u_p)^{3/2}$, leading to Eq.~(3.41) of Ref.~\cite{Hinderer:2017jcs}, which reads
\begin{equation}
\mathrm{e}^{-ik\psi_r} = \mathrm{e}^{ik \frac{e\sqrt{1-e^2}\sin\chi}{1+e\cos\chi}} \left(\frac{1+\sqrt{1-e^2} + e\,\mathrm{e}^{i\chi}}{e + (1 + \sqrt{1 - e^2})\mathrm{e}^{i\chi}}\right)^k.
\end{equation}

Hence, the modes with $\ell = 2$ and $m$ even are given by
\begin{align}
h_\text{tail}^{2m} &= \frac{\sqrt{24} M a_{2m}}{c^7 R} \mathrm{e}^{-im\phi} 
\sum_{k=-\infty}^{\infty} J_{2mk} \Omega_{mk}^4  \nonumber\\
&\quad\qquad \times \mathrm{e}^{im\chi} \mathrm{e}^{-i(k + m)\psi_r} \mathcal{I}(\Omega_{mk}),
\end{align}
while the modes with $\ell = 3$ and $m$ odd are given by
\begin{align}
h_\text{tail}^{3m} &= -i \frac{4\sqrt{5} M a_{3m}}{3\sqrt{6}c^8 R} \mathrm{e}^{-im\phi} 
\sum_{k=-\infty}^{\infty} J_{3mk} \Omega_{mk}^5 \nonumber\\
&\quad\qquad \times \mathrm{e}^{im\chi} \mathrm{e}^{-i(k + m)\psi_r} \mathcal{I}(\Omega_{mk}).
\end{align}

To obtain analytical expressions for the modes, we expand the above equations in eccentricity, where the infinite sum over $k$ can be stopped at the order of the expansion in $e$.
The result of that expansion is complicated, but we can perform a phase redefinition in the leading order instantaneous part\footnote{
One first needs to express the leading order part in terms of the variables ($e,x,\chi$) instead of ($r,p_r,\pphi$) using the relations from Appendix~\ref{app:kepler}. For example, for the $(2,2)$ mode, we obtain
\begin{equation*}
h_\text{LO}^{22} = \frac{-8\mu}{c^4D_L} \sqrt{\frac{\pi}{5}} \mathrm{e}^{-2i\phi}\frac{x}{1 - e^2} \left[1 + \frac{e}{4} (\mathrm{e}^{-i\chi} + 5\mathrm{e}^{i\chi}) + \frac{e^2}{2} \mathrm{e}^{2i\chi} \right].
\end{equation*}
} of the form $\phi \to \phi + x^{3/2}\delta_\phi$, and absorb in $\delta_\phi$ all terms that are not proportional to $\pi^{3/2}$. This modifies the phase at 4PN relative order, which we can ignore when working to 2PN order. (See Ref.~\cite{Kidder:2007rt} for more details.) The result for the $(2,2)$ mode to $\Order(e^6)$ is given by
\begin{widetext}
\begin{align}
\hat{H}_\text{tail}^{22} &= \frac{2\pi}{c^3} x^{5/2} \bigg[
1 + e \left(\frac{11 \mathrm{e}^{-i \chi }}{8}+\frac{13 \mathrm{e}^{i \chi }}{8}\right) 
+ e^2 \left(\frac{5}{8} \mathrm{e}^{-2 i \chi }+\frac{7}{8} \mathrm{e}^{2 i \chi }+4\right)
+e^3 \left(\frac{121 \mathrm{e}^{-i \chi }}{32}+\frac{143 \mathrm{e}^{i \chi }}{32}+\frac{3}{32} \mathrm{e}^{-3 i \chi }+\frac{1}{12} \mathrm{e}^{3 i \chi }\right) \nonumber\\
&\quad
+e^4 \left(\frac{25}{16} \mathrm{e}^{-2 i \chi }+\frac{203}{96} \mathrm{e}^{2 i \chi }-\frac{5}{96} \mathrm{e}^{4 i \chi }+\frac{65}{8}\right) 
+e^5 \left(\frac{55 \mathrm{e}^{-i \chi }}{8}+\frac{6233 \mathrm{e}^{i \chi }}{768}+\frac{15}{64} \mathrm{e}^{-3 i \chi }+\frac{281 \mathrm{e}^{3 i \chi }}{1536}+\frac{53 \mathrm{e}^{5 i \chi }}{7680}\right) \nonumber\\
&\quad
+e^6 \left(\frac{175}{64} \mathrm{e}^{-2 i \chi }+\frac{1869}{512} \mathrm{e}^{2 i \chi }-\frac{449 \mathrm{e}^{4 i \chi }}{3840}+\frac{31 \mathrm{e}^{6 i \chi }}{23040}+\frac{30247}{2304}\right)
\bigg],
\end{align}
while for the $(2,0)$ mode
\begin{align}
\hat{H}_\text{tail}^{20} &=  \frac{\pi x^{5/2}}{2\sqrt{6}c^3} \bigg[
e \left( \mathrm{e}^{-i \chi } + \mathrm{e}^{i \chi }\right) +e^2 \left(\mathrm{e}^{-2 i \chi } + \mathrm{e}^{2 i \chi }+2\right) 
+e^3 \left(3  \mathrm{e}^{-i \chi } + 3  \mathrm{e}^{i \chi }+\frac{1}{4} \mathrm{e}^{-3 i \chi }+\frac{1}{4} \mathrm{e}^{3 i \chi }\right) \nonumber\\
&\quad
+e^4 \left(\frac{29}{12} \mathrm{e}^{-2 i \chi }+\frac{29}{12} \mathrm{e}^{2 i \chi }+\frac{9}{2}\right) 
+e^5 \left(\frac{179 \mathrm{e}^{-i \chi }}{32}+\frac{179 \mathrm{e}^{i \chi }}{32}+\frac{125}{192} \mathrm{e}^{-3 i \chi }+\frac{125}{192} \mathrm{e}^{3 i \chi }\right)  \nonumber\\
&\quad
+e^6 \left(\frac{805}{192} \mathrm{e}^{-2 i \chi }+\frac{805}{192} \mathrm{e}^{2 i \chi }-\frac{7}{960} \mathrm{e}^{-4 i \chi }-\frac{7}{960} \mathrm{e}^{4 i \chi }+\frac{121}{16}\right)
\bigg].
\end{align}
The $(3,3)$ mode is given by
\begin{align}
\hat{H}_\text{tail}^{33} &= -\frac{9i \pi \delta}{4c^4}\sqrt{\frac{15}{14}} x^3  \bigg[
1+e \left(\frac{47 \mathrm{e}^{-i \chi }}{27}+\frac{19 \mathrm{e}^{i \chi }}{9}\right) 
+e^2 \left(\frac{61}{54} \mathrm{e}^{-2 i \chi }+\frac{91}{54} \mathrm{e}^{2 i \chi }+\frac{155}{27}\right) 
+e^3 \bigg(\frac{691 \mathrm{e}^{-i \chi }}{108}+\frac{841 \mathrm{e}^{i \chi }}{108}+\frac{35\mathrm{e}^{-3 i \chi }}{108}  \nonumber\\
&\quad\qquad 
+\frac{65\mathrm{e}^{3 i \chi }}{108} \bigg) 
+e^4 \left(\frac{32}{9} \mathrm{e}^{-2 i \chi }+\frac{287}{54} \mathrm{e}^{2 i \chi }+\frac{5 \mathrm{e}^{-4 i \chi }}{144}+\frac{115 \mathrm{e}^{4 i \chi }}{1728}+\frac{3139}{216}\right) 
+e^5 \bigg(\frac{503 \mathrm{e}^{-i \chi }}{36}+\frac{613 \mathrm{e}^{i \chi }}{36} +\frac{35\mathrm{e}^{-3 i \chi }}{36}  \nonumber\\
&\quad\qquad 
+\frac{3095 \mathrm{e}^{3 i \chi }}{1728}-\frac{457 \mathrm{e}^{5 i \chi }}{25920}\bigg) 
+e^6 \left(\frac{131}{18}\mathrm{e}^{-2 i \chi }+\frac{150503 \mathrm{e}^{2 i \chi }}{13824}+\frac{5}{48} \mathrm{e}^{-4 i \chi }+\frac{151}{810} \mathrm{e}^{4 i \chi }-\frac{41 \mathrm{e}^{6 i \chi }}{20736}+\frac{219}{8}\right)
\bigg],
\end{align}
and the $(3,1)$ mode
\begin{align}
\hat{H}_\text{tail}^{31} &= \frac{i\pi \delta x^3}{12\sqrt{14}c^4} \bigg[
1+e \left(\mathrm{e}^{i \chi }-9 \mathrm{e}^{-i \chi }\right) 
+ e^2 \left(-\frac{27}{2} \mathrm{e}^{-2 i \chi }-\frac{5}{4} \mathrm{e}^{2 i \chi }-15\right) 
+e^3 \left(-\frac{177}{4} \mathrm{e}^{-i \chi }-\frac{19 \mathrm{e}^{i \chi }}{2}-\frac{25}{4} \mathrm{e}^{-3 i \chi }-\frac{4}{3} e^{3 i \chi }\right) \nonumber\\
&\quad
+e^4 \left(-\frac{89}{2} \mathrm{e}^{-2 i \chi }-\frac{125}{24} \mathrm{e}^{2 i \chi }-\frac{15}{16} \mathrm{e}^{-4 i \chi }-\frac{55\mathrm{e}^{4 i \chi }}{192} -\frac{1703}{32}\right)
+e^5 \bigg(-\frac{10141}{96} \mathrm{e}^{-i \chi }-\frac{2867 \mathrm{e}^{i \chi }}{96}-\frac{75}{4} \mathrm{e}^{-3 i \chi }-\frac{629}{192} \mathrm{e}^{3 i \chi } \nonumber\\
&\quad\qquad
-\frac{7}{960} \mathrm{e}^{5 i \chi }\bigg)
+e^6 \left(-\frac{142903 \mathrm{e}^{-2 i \chi }}{1536}-\frac{2965}{256} \mathrm{e}^{2 i \chi }-\frac{45}{16} \mathrm{e}^{-4 i \chi }-\frac{239}{240} \mathrm{e}^{4 i \chi }+\frac{37 \mathrm{e}^{6 i \chi }}{23040}-\frac{16343}{144}\right)
\bigg].
\end{align}
We checked that our results agree with those of Ref.~\cite{Boetzel:2019nfw} after converting between the quasi-Keplerian and Keplerian parametrization, and performing a phase shift.

To express the modes in terms of $(r,p_r,\pphi)$ instead of $(x,e,\chi)$, we use the following leading order relations:
\begin{equation}
\pphi = \frac{1}{\sqrt{u_p}}, \qquad p_r = e \sqrt{u_p} \sin\chi, \qquad
\frac{1}{r} = u_p (1 + e \cos\chi), \qquad  x = u_p (1 - e^2).
\end{equation}
As explained above, it is advantageous to replace $\pphi^2$ with $\dot{p}_r$ using $\dot{p}_r = (\pphi^2 - r)/r^3$ and expand in both $p_r$ and $\dot{p}_r$  (since $p_r$ and $\dot{p}_r$ are both of order $e$) to obtain
\begin{align}
\hat{H}_\text{tail}^{22} &= \frac{2\pi}{c^3} \Bigg\lbrace 
\frac{\pphi}{r^3}+\frac{i p_r}{4 r^2} 
+\left[\frac{7}{32} \pphi p_r^2 \dot{p}_r-\frac{7}{96}  \pphi r^3 \dot{p}_r^3+i \left(\frac{7 p_r^3}{96 r}-\frac{7}{32} r^2 p_r \dot{p}_r^2\right)\right]
+ \bigg[\frac{3\pphi}{32}  r^5 \dot{p}_r^4-\frac{\pphi}{8}  r^2 p_r^2 \dot{p}_r^2+\frac{\pphi p_r^4}{48 r} \nonumber\\
&\quad +i \left(\frac{ r^4}{12} p_r \dot{p}_r^3-\frac{ r}{96} p_r^3 \dot{p}_r\right)\bigg]  + \left[\frac{31}{384} \pphi r p_r^4 \dot{p}_r-\frac{173 \pphi r^7 \dot{p}_r^5}{1920}-\frac{1}{192} \pphi r^4 p_r^2 \dot{p}_r^3+i \left(\frac{r^6 p_r}{768}  \dot{p}_r^4-\frac{49}{384} r^3 p_r^3 \dot{p}_r^2+\frac{89 p_r^5}{3840}\right)\right] \nonumber\\
&\quad + \Bigg[\frac{97 \pphi r^9 \dot{p}_r^6}{1152}+\frac{1}{16} \pphi r^6 p_r^2 \dot{p}_r^4 
-\frac{47}{384} \pphi r^3 p_r^4 \dot{p}_r^2+\frac{\pphi p_r^6}{96}+i \left(-\frac{1}{64} r^8 p_r \dot{p}_r^5+\frac{137 r^5 p_r^3 \dot{p}_r^3}{1152}-\frac{23}{640} r^2 p_r^5 \dot{p}_r\right)\Bigg]
\Bigg\rbrace, \nonumber\\
\hat{H}_\text{tail}^{20} &= \frac{\pi}{\sqrt{6}c^3} \Bigg\lbrace
\frac{\pphi  \dot{p}_r}{r} - \pphi r\dot{p}_r^2
+ \left[\frac{3}{4} \pphi r^3 \dot{p}_r^3-\frac{1}{4} \pphi p_r^2 \dot{p}_r\right]
+ \left[-\frac{7}{12} \pphi r^5 \dot{p}_r^4-\frac{\pphi p_r^4}{6 r}\right]
+ \bigg[\frac{95}{192} \pphi r^7 \dot{p}_r^5+\frac{13}{96} \pphi r^4 p_r^2 \dot{p}_r^3 \nonumber\\
&\quad
+\frac{11}{192} \pphi r p_r^4 \dot{p}_r\bigg]
+ \left[-\frac{139}{320} \pphi r^9 \dot{p}_r^6-\frac{11}{96} \pphi r^6 p_r^2 \dot{p}_r^4+\frac{3}{64} \pphi r^3 p_r^4 \dot{p}_r^2-\frac{\pphi p_r^6}{480}\right]
\Bigg\rbrace, \nonumber\\
\hat{H}_\text{tail}^{33} &= -\frac{9i \pi \delta}{4c^4}\sqrt{\frac{15}{14}} \Bigg\lbrace
\frac{1}{r^3} + \left[\frac{23 \dot{p}_r}{27 r}+\frac{10 i \pphi p_r}{27 r^3}\right]-\frac{2 p_r^2}{27 r^2}
+ \bigg[i \left(\frac{25}{432} \pphi p_r^3 \dot{p}_r-\frac{25}{432} \pphi r^3 p_r \dot{p}_r^3\right)-\frac{25 r^5 \dot{p}_r^4}{1728}+\frac{25}{288} r^2 p_r^2 \dot{p}_r^2 \nonumber\\
&\quad
-\frac{25 p_r^4}{1728 r}\bigg]
+ \left[i \left(\frac{109 \pphi r^5 p_r \dot{p}_r^4}{2592}+\frac{41 \pphi r^2 p_r^3 \dot{p}_r^2}{1296}-\frac{41 \pphi p_r^5}{12960 r}\right)+\frac{293 r^7 \dot{p}_r^5}{25920}+\frac{41 r^4 p_r^2 \dot{p}_r^3}{1296}-\frac{157 r p_r^4 \dot{p}_r}{5184}\right] + \bigg[\frac{2845 r^3 p_r^4 \dot{p}_r^2}{41472} \nonumber\\
&\quad
-\frac{1561 r^9 \dot{p}_r^6}{207360}-\frac{775 r^6 p_r^2 \dot{p}_r^4}{13824}-\frac{653 p_r^6}{69120}
+i \left(\frac{4211 \pphi r p_r^5 \dot{p}_r}{103680}-\frac{2173 \pphi r^7 p_r \dot{p}_r^5}{103680}-\frac{307 \pphi r^4 p_r^3 \dot{p}_r^3}{3456}\right)\bigg]
\Bigg\rbrace, \nonumber\\
\hat{H}_\text{tail}^{31} &= \frac{i\pi \delta}{12\sqrt{14}c^4} \Bigg\lbrace
\frac{1}{r^3}+  \left[\frac{10 i \pphi p_r}{r^3}-\frac{11 \dot{p}_r}{r}\right]
- \left[\frac{11 i \pphi \dot{p}_r p_r}{2 r} + \frac{13 p_r^2}{4 r^2} + \frac{11 r \dot{p}_r^2}{4} \right]
+ \bigg[i \left(6 \pphi r p_r \dot{p}_r^2-\frac{\pphi p_r^3}{6 r^2}\right)+\frac{35}{12} r^3 \dot{p}_r^3 \nonumber\\
&\quad
 -\frac{p_r^2 \dot{p}_r}{2} \bigg]
+ \left[i \left(\frac{45}{16} \pphi p_r^3 \dot{p}_r-\frac{61}{16} \pphi r^3 p_r \dot{p}_r^3\right)-\frac{269}{192} r^5 \dot{p}_r^4+\frac{37}{32} r^2 p_r^2 \dot{p}_r^2-\frac{63 p_r^4}{64 r}\right] 
+ \Bigg[
\frac{643}{960} r^7 \dot{p}_r^5 +\frac{5}{16} r^4 p_r^2 \dot{p}_r^3 \nonumber\\
&\quad
+\frac{101}{192} r p_r^4 \dot{p}_r
+i \left(\frac{83}{32} \pphi r^5 p_r \dot{p}_r^4-\frac{43}{48} \pphi r^2 p_r^3 \dot{p}_r^2+\frac{379 \pphi p_r^5}{480 r}\right)
\Bigg]
+ \Bigg[
\frac{317}{768} r^3 p_r^4 \dot{p}_r^2  -\frac{5039 r^9 \dot{p}_r^6}{11520}-\frac{289}{768} r^6 p_r^2 \dot{p}_r^4-\frac{297 p_r^6}{1280}
\nonumber\\
&\quad
+i \left(\frac{151}{576} \pphi r^4 p_r^3 \dot{p}_r^3-\frac{4267 \pphi r^7 p_r \dot{p}_r^5}{1920}+\frac{271}{640} \pphi r p_r^5 \dot{p}_r\right)\Bigg]
\Bigg\rbrace.
\end{align} 
\end{widetext}

\subsubsection{Modes with odd $\ell + m$}
The Newtonian order current quadrupole moment is given by~\cite{Arun:2007rg}
\begin{equation}
J^{ij} = -\delta \mu r^2 n_k v_l \epsilon^{kl\langle i} n^{j\rangle} = -\delta r \pphi \, \hat{e}_z^{\langle i} n^{j\rangle},
\end{equation}
where $\hat{e}_z^i$ is the unit vector in the $z$-direction. The term $e_z^{\langle i} n^{j\rangle}$ can be expressed in terms of $\mathcal{Y}^{ij}_{21}$, as was done in Ref.~\cite{Banihashemi:2018xfb}, by defining the complex vector 
\begin{equation}
\zeta^i = e_x^i + i e_y^i,
\end{equation}
which leads to
\begin{equation}
\mathcal{Y}_{21}^{ij} = -\frac{1}{2} \sqrt{\frac{15}{2\pi}} \zeta^{\langle i} e_z^{j\rangle}.
\end{equation}
Since, for equatorial orbits, $n^i = \cos\phi \,\hat{e}_x^i + \sin\phi \,\hat{e}_y^i$ and ${\lambda^i = -\sin\phi \,\hat{e}_x^i + \cos\phi \,\hat{e}_y^i}$, we obtain
\begin{equation}
n^i + i\lambda^i = e^{-i\phi} \zeta^i.
\end{equation}
Hence,
\begin{align}
e_z^{\langle i} n^{j\rangle} &= \text{Re}\left[\mathrm{e}^{-i\phi} e_z^{\langle i} \zeta^{j\rangle} \right] = -2 \sqrt{\frac{2\pi}{15}} \text{Re}\left[\mathrm{e}^{-i\phi}\mathcal{Y}_{21}^{ij}\right] \nonumber\\
&= -\sqrt{\frac{2\pi}{15}} \left(\mathrm{e}^{-i\phi}\mathcal{Y}_{21}^{ij} + \mathrm{e}^{i\phi}\bar{\mathcal{Y}}_{21}^{ij}\right).
\end{align}
Since $V_{ij}$ is contracted with $\bar{\mathcal{Y}}_{ij}^{\ell m}$ in Eq.~\eqref{Vlm}, and  $\bar{\mathcal{Y}}_{ij}^{\ell m}\bar{\mathcal{Y}}^{ij}_{\ell m}=0$, only the term with $\mathcal{Y}_{21}^{ij}$ in the above equation contributes to the modes. Thus, we only need to consider the following part of the current quadrupole
\begin{equation}
J^{ij} = \mu \delta \sqrt{\frac{2\pi}{15}}\mathcal{Y}_{21}^{ij} \pphi r e^{-i\phi} + \dots.
\end{equation}

Then, we follow the same steps as in the previous subsection. Decomposing the phase into $\phi = \omega_\phi t + \Delta\phi$ leads to
\begin{equation}
J^{ij} =  \mu \delta \sqrt{\frac{2\pi}{15}} \mathcal{Y}_{21}^{ij} \pphi \mathrm{e}^{-i\psi_\phi} J_{11} + \dots,
\end{equation}
with
\begin{equation}
J_{11} = r \mathrm{e}^{-i\Delta\phi} = \sum_{k=-\infty}^{\infty} J_{11k} \mathrm{e}^{-ik\psi_r},
\end{equation}
and
\begin{align}
J_{11k} &= \frac{1}{2\pi} \int_{0}^{2\pi} d\psi_r \, J_{11}\mathrm{e}^{ik\psi_r} \nonumber\\
&=\frac{\omega_r}{2\pi u_p^{5/2}} \int_{0}^{2\pi} \frac{d\chi}{(1+e\cos\chi)^3} \mathrm{e}^{-i\Delta \phi} \mathrm{e}^{ik\psi_r}.
\end{align}
Thus, the current quadrupole source moment can be expressed as
\begin{equation}
J^{ij} = \mu \delta \sqrt{\frac{2\pi}{15}}\mathcal{Y}_{21}^{ij} \pphi \sum_{k=-\infty}^{\infty} J_{11k} \mathrm{e}^{-i(k\psi_r+\psi_\phi)} + \dots,
\end{equation}
and the current quadrupole radiative moment
\begin{align}
V_{ij}^\text{tail} &= \frac{2M\mu\delta}{c^3}\sqrt{\frac{2\pi}{15}} \mathcal{Y}_{21}^{ij} \pphi  \sum_{k=-\infty}^{\infty} J_{11k} \Omega_{1k}^4 \nonumber\\
&\quad\qquad
\times \mathrm{e}^{-i(k\psi_r + \psi_\phi)} \mathcal{I}(\Omega_{1k}),
\end{align}
leading to the $(2,1)$ mode
\begin{align}
h_\text{tail}^{21} &= \frac{16 i}{3} \sqrt{\frac{\pi}{5}} \frac{\delta M}{Rc^8}\mathrm{e}^{-i\phi} \pphi \sum_{k=-\infty}^{\infty} J_{11k} \Omega_{1k}^4 \mathrm{e}^{i\chi} \nonumber\\
&\quad\qquad
\times\mathrm{e}^{-i(k + 1)\psi_r} \mathcal{I}(\Omega_{1k}),
\end{align}
with $\pphi = 1/\sqrt{u_p}=\sqrt{(1-e^2)/x}$.
Expanding in eccentricity yields
\begin{widetext}
\begin{align}
\hat{H}_\text{tail}^{21} &= \frac{i\pi \delta}{3c^4}x^3 \bigg[
1+e \left(3 \mathrm{e}^{-i \chi }+\mathrm{e}^{i \chi }\right) 
+ e^2 \left(3 \mathrm{e}^{-2 i \chi }+\frac{1}{4} \mathrm{e}^{2 i \chi }+6\right) 
+e^3 \left(\frac{45 \mathrm{e}^{-i \chi }}{4}+4 \mathrm{e}^{i \chi }+\frac{5}{4} \mathrm{e}^{-3 i \chi }+\frac{1}{6} \mathrm{e}^{3 i \chi }\right) \nonumber\\
&\quad
+ e^4 \left(\frac{19}{2} \mathrm{e}^{-2 i \chi }+\frac{25}{24} \mathrm{e}^{2 i \chi }+\frac{3}{16} \mathrm{e}^{-4 i \chi }+\frac{17}{192} \mathrm{e}^{4 i \chi }+\frac{493}{32}\right) 
+ e^5 \bigg(\frac{2375 \mathrm{e}^{-i \chi }}{96}+\frac{865 \mathrm{e}^{i \chi }}{96}+\frac{15}{4} \mathrm{e}^{-3 i \chi }+\frac{91}{192} \mathrm{e}^{3 i \chi } \nonumber\\
&\quad\qquad 
-\frac{7}{960} \mathrm{e}^{5 i \chi }\bigg) 
+e^6 \left(\frac{29957 \mathrm{e}^{-2 i \chi }}{1536}+\frac{593}{256} \mathrm{e}^{2 i \chi }+\frac{9}{16} \mathrm{e}^{-4 i \chi }+\frac{241}{960} \mathrm{e}^{4 i \chi }+\frac{37 \mathrm{e}^{6 i \chi }}{23040}+\frac{8417}{288}\right)
\bigg],
\end{align}
which is in agreement with the results of Ref.~\cite{Boetzel:2019nfw}.
In terms of $(r,p_r,\pphi,\dot{p}_r)$, we obtain
\begin{align}
\hat{H}_\text{tail}^{21} &= \frac{i\pi \delta}{3c^4} \Bigg\lbrace
\frac{1}{r^3}+ \left[\frac{\dot{p}_r}{r}-\frac{2 i \pphi p_r}{r^3}\right]
+ \left[\frac{i \pphi \dot{p}_r p_r}{2 r}-\frac{p_r^2}{4 r^2}+\frac{1}{4} r \dot{p}_r^2\right]
+ \left[-\frac{i \pphi p_r^3}{6 r^2}-\frac{1}{12} r^3 \dot{p}_r^3-\frac{1}{2} \dot{p}_r p_r^2\right] \nonumber\\
&\quad
+\Bigg[i \left(\frac{5}{16} \pphi p_r^3 \dot{p}_r-\frac{5}{16} \pphi r^3 p_r \dot{p}_r^3\right)
-\frac{1}{192} 5 r^5 \dot{p}_r^4+\frac{13}{32} r^2 p_r^2 \dot{p}_r^2-\frac{7 p_r^4}{64 r}\Bigg]
+ \Bigg[i \left(\frac{11}{32} \pphi r^5 p_r \dot{p}_r^4-\frac{19}{48} \pphi r^2 p_r^3 \dot{p}_r^2+\frac{19 \pphi p_r^5}{480 r}\right) \nonumber\\
&\quad
+\frac{43}{960} r^7 \dot{p}_r^5-\frac{3}{16} r^4 p_r^2 \dot{p}_r^3+\frac{29}{192} r p_r^4 \dot{p}_r\Bigg] 
+\bigg[i \left(-\frac{511 \pphi r^7 p_r \dot{p}_r^5}{1920}+\frac{115}{576} \pphi r^4 p_r^3 \dot{p}_r^3-\frac{61}{640} \pphi r p_r^5 \dot{p}_r\right)
-\frac{79 r^9 \dot{p}_r^6}{2304} \nonumber\\
&\quad 
-\frac{13}{768} r^6 p_r^2 \dot{p}_r^4 +\frac{17}{768} r^3 p_r^4 \dot{p}_r^2+\frac{59 p_r^6}{1280}\bigg]\Bigg\rbrace.
\end{align}
\end{widetext}

\subsection{Aligned-spin contributions}
The spin contributions to the modes were derived for circular orbits in Refs.~\cite{Buonanno:2012rv,Arun:2008kb,Siemonsen:2017yux}. 
To derive the spin part of the modes to 2PN for eccentric orbits, we use the source moments from Refs.~\cite{Blanchet:2006gy,Buonanno:2012rv}, which are in harmonic coordinates and in terms of the covariant SSC.
Differentiating the source moments to obtain the radiative moments~\eqref{radMoments}, and plugging them into Eq.~\eqref{hlm}, we obtain the modes listed in Appendix~\ref{app:harmModes}.
Transforming from harmonic to EOB coordinates, and from the covariant to the NW SSC using the transformations in Appendix~\ref{app:transform}, we obtain the following spin contributions to the modes:
\begin{widetext}
\begin{align}
\label{spinmodes}
\hat{H}_\text{spin}^{22} &= \frac{1}{c^3} \left[\frac{\chi _1 }{12 r^3} \left[(6 \delta +\nu +6) p_{\phi }+2 i (3 \delta -\nu +3) r p_r\right]
+\frac{\chi _2}{12 r^3}\left[(-6 \delta +\nu +6) p_{\phi }-2 i (3 \delta +\nu -3) r p_r\right]\right] \nonumber\\
&\quad  + \frac{1}{4c^4 r^3} \left\lbrace
\chi _1^2 \left[3 \CESA X_1^2-X_1^4 \left(2 p^2 r+1\right)\right]
+\chi _2^2 \left[3 \CESB X_2^2-X_2^4 \left(2 p^2 r+1\right)\right]
-2 \nu  \chi _1 \chi _2 \left(\nu-3 +2 \nu  p^2 r\right)
\right\rbrace,\nonumber\\
\hat{H}_\text{spin}^{21} &= \frac{i}{4c^2 r^2} \left[-(1 + \delta) \chi_1 + (1-\delta) \chi_2\right]
+ \frac{\chi_1}{84 c^4 r^2} 
\bigg[\frac{i p_{\phi }^2}{r^2} (43 \nu\delta -42\delta+153 \nu -42) +\frac{ p_r p_{\phi }}{r}(3 \delta  (2 \nu +49)-104 \nu +147) \nonumber\\
&\quad +\frac{i}{2}  p_r^2(\delta  (38 \nu +105)+74 \nu +105) +\frac{i }{r}(  63\delta-38 \nu\delta-74 \nu +63)\bigg]
+ \frac{\chi_2}{84 c^4 r^2} 
\bigg[\frac{i  p_{\phi }^2}{r^2}(43 \nu\delta -42\delta-153 \nu +42) \nonumber\\
&\quad
+\frac{p_r p_{\phi }}{r}(3 \delta  (2 \nu +49)+104 \nu -147) 
+\frac{i}{2} p_r^2 (38 \nu\delta +105\delta-74 \nu -105) +\frac{i }{r}(  63\delta-38 \nu\delta+74 \nu -63)\bigg], \nonumber\\
\hat{H}_\text{spin}^{20} &= \frac{\sqrt{3}\pphi}{2\sqrt{2}c^3r^3} \left[ (2\delta -\nu +2) \chi_1 + (-2 \delta -\nu + 2) \chi_2\right] 
+ \frac{\sqrt{3}}{2\sqrt{2}c^4r^3} \bigg\lbrace 
\chi _1^2 \left[\frac{X_1^4 }{3 r}\left(2 r^2 p_r^2-2 p_{\phi }^2+r\right)-\CESA X_1^2\right] \nonumber\\
&\quad\qquad 
+ \chi _2^2 \left[\frac{X_2^4}{3 r} \left(2 r^2 p_r^2-2 p_{\phi }^2+r\right)-\CESB X_2^2\right]
+\frac{2 \nu  \chi _1 \chi _2 }{3} \bigg[\nu -3-2 \nu  \frac{p_{\phi }^2}{r}+2 \nu  r p_r^2 \bigg]
\bigg\rbrace,\nonumber \\
\hat{H}_\text{spin}^{30} &= \frac{-i\nu p_r}{\sqrt{42} c^3 r^2} (\chi_1 + \chi_2), \nonumber\\
\hat{H}_\text{spin}^{31} &= \frac{1}{24 \sqrt{14} c^4 r^2} \Bigg\lbrace
\chi_1  \bigg[\frac{i  p_{\phi }^2}{2 r^2} ( 55 \nu\delta -96\delta+375 \nu -96)  +i\left(p_r^2-\frac{2 }{r}\right) ( 2 \nu\delta -6 \delta+23 \nu -6) \nonumber\\
&\quad 
+\frac{ p_r p_{\phi }}{r} ( -6 \nu\delta + 30 \delta-127 \nu +30)\bigg] 
+\chi_2\bigg[
\frac{i  p_{\phi }^2}{2 r^2}(55 \nu\delta -96\delta -375 \nu +96)+\frac{ p_r p_{\phi }}{r}(-6\nu\delta + 30 \delta +127 \nu -30) \nonumber\\
&\quad 
+i\left(p_r^2 -\frac{2}{r}\right) (2 \nu\delta - 6 \delta-23 \nu +6)\bigg] 
\Bigg\rbrace, \nonumber\\
\hat{H}_\text{spin}^{32} &= \frac{\nu}{6c^3 r^3}\sqrt{\frac{5}{7}} (4 \pphi + i r p_r) (\chi_1 + \chi_2), \nonumber\\
\hat{H}_\text{spin}^{33} &= \frac{\sqrt{5}}{8 \sqrt{42} c^4 r^2} \Bigg\lbrace
\left[-\frac{23 i (\delta +1) \nu  p_{\phi }^2}{2 r^2}+\frac{(2 \nu\delta + 6 \delta-19 \nu +6) p_r p_{\phi }}{r} + i\left(\frac{2}{r}-p_r^2\right) (2 \nu\delta - 6 \delta +23 \nu -6)\right] \chi_1 \nonumber\\
&\quad
+\left[-\frac{23 i (\delta -1) \nu  p_{\phi }^2}{2 r^2}+\frac{( 2 \delta\nu +6 \delta +19 \nu -6) p_r p_{\phi }}{r}+ i\left(\frac{2}{r}-p_r^2\right)(2 \delta\nu - 6 \delta -23 \nu +6)\right] \chi_2
\Bigg\rbrace, \nonumber\\
\hat{H}_\text{spin}^{41} &= - i \sqrt{\frac{5}{2}} \frac{\nu}{336 c^4 r^4}  \left(-10 i r p_r p_{\phi }+ 6 r^2 p_r^2-12 r+11 p_{\phi }^2\right) \left[(\delta -1) \chi _1+(\delta +1) \chi _2\right], \nonumber \\
\hat{H}_\text{spin}^{43} &=  \sqrt{\frac{5}{14}} \frac{\nu}{48 c^4 r^4}  \left(10 r p_r p_{\phi }+ 2 i r^2 p_r^2- 4 i r-23 i p_{\phi }^2\right) \left[(\delta -1) \chi _1+(\delta +1) \chi _2\right].
\end{align}
\end{widetext}
The circular-orbit limit of these modes, when expressed in terms of the orbital frequency, agrees with the results of Refs.~\cite{Buonanno:2012rv,Siemonsen:2017yux}.
The spin contributions to the $(2,2),~(2,1),$ and $(3,3)$ modes for eccentric orbits were calculated in Ref.~\cite{Liu:2021pkr}; however, we find a small disagreement with their results for the SO part.\footnote{
The difference between the modes in Ref.~\cite{Liu:2021pkr} (denoted with a bar) and the modes in Eq.~\eqref{spinmodes} (with $\CESA = \CESB=1$) is given by
\begin{align*}
\hat{\bar{H}}_\text{spin}^{22} - \hat{H}_\text{spin}^{22} &= \frac{i \nu   p_r}{2 c^3 r^2} \left(\chi _1+\chi _2\right), \\
\hat{\bar{H}}_\text{spin}^{21} - \hat{H}_\text{spin}^{21} &= \frac{i\delta\nu p_r}{6c^4 r^3}  \left(r p_r+i p_{\phi }\right) \left(\chi _1+\chi _2\right),\\
\hat{\bar{H}}_\text{spin}^{33} - \hat{H}_\text{spin}^{33} &= \frac{\sqrt{5} \delta  \nu  p_r }{8 \sqrt{42} c^4 r^3} \left(17 p_{\phi }+5 i r p_r\right) \left(\chi _1+\chi _2\right),
\end{align*}
which is likely due to the coordinate/SSC transformations detailed in Appendix~\ref{app:transform}.
}

\subsection{Factorized modes}
\label{sec:factModes}

The quasi-circular waveform modes used in \texttt{SEOBNRv4HM} are factorized as follows~\cite{Damour:2007yf,Damour:2008gu,Pan:2010hz,Cotesta:2018fcv}:
\begin{equation}
h_{\ell m}^\text{F,qc} = h_{\ell m}^\text{N,qc} \hat{S}_\text{eff}^\text{qc} T_{\ell m}^\text{qc} e^{i\delta_{\ell m}} f_{\ell m}^\text{qc},
\end{equation}
where $h_{\ell m}^\text{N,qc}$ is the Newtonian part of the mode, $\hat{S}_\text{eff}^\text{qc}$ is an effective source term given by
\begin{equation}
\hat{S}_\text{eff}^\text{qc} = \left\{
        \begin{array}{ll}
            \hat{H}_\text{eff}(v_\Omega) & \quad \ell + m \text{ even} \\
            \hat{L}_\text{eff} \equiv v_\Omega\, \pphi(v_\Omega) & \quad \ell + m \text{ odd}
        \end{array}
    \right. ,
\end{equation}
$T_{\ell m}^\text{qc}$ resums the infinite number of ``leading logarithms'' entering the tail effects, $\delta_{\ell m}$ contains the part of the tail not included in $T_{\ell m}^\text{qc}$, and $f_{\ell m}$ contains PN corrections such that the expansion of $h_{\ell m}^\text{F,qc}$ agrees with the known PN expansion of the modes. See Refs.~\cite{Cotesta:2018fcv,Pan:2010hz} for more details and for expressions of these terms.

We include the eccentric corrections in the factorized modes as follows:
\begin{align}
\label{fEmodes}
h_{\ell 0}^\text{F} &= \hat{S}_\text{eff} (1 + T_{\ell 0}^\text{ecc}) f_{\ell 0}^\text{ecc}, \nonumber\\
h_{\ell m}^\text{F} &= h_{\ell m}^\text{N,qc} \hat{S}_\text{eff} (T_{\ell m}^\text{qc} + T_{\ell m}^\text{ecc}) e^{i\delta_{\ell m}} (f_{\ell m}^\text{qc} + f_{\ell m}^\text{ecc}),
\end{align}
where the effective source term is given by
\begin{equation}
\hat{S}_\text{eff} = \left\{
        \begin{array}{ll}
            \hat{H}_\text{eff}(r,p_r,\pphi) & \quad \ell + m \text{ even} \\
            \hat{L}_\text{eff} \equiv v_\Omega\, \pphi & \quad \ell + m \text{ odd}
        \end{array}
    \right. ,
\end{equation}
$T_{\ell m}^\text{ecc}$ contains the eccentric corrections to the hereditary contributions, $\delta_{\ell m}$ is the same as in the quasi-circular case, and $f_{\ell m}^\text{ecc}$ contains the eccentric corrections to the instantaneous contributions (both spinning and nonspinning, including the Newtonian part).
For example, for the leading order of the $(2,2)$ mode, we obtain
\begin{align}
f_{22}^\text{ecc} &= \frac{1}{2 (r^2 \dot{p}_r+1)^{1/3}} \bigg[
2 + r^2 \dot{p}_r  -r p_r^2   -2 \left(r^2 \dot{p}_r+1\right)^{1/3} \nonumber\\
&\quad\qquad+2 i  p_r\sqrt{r^3 \dot{p}_r+r}
\bigg] + \dots.
\end{align}

For the tail part, we simplified the results of Sec.~\ref{sec:modesTail} and eliminated the gauge parameter by using a phase shift, which led to the circular part of the tail contribution to the $(2,2)$ mode simply being $2\pi v_\Omega^5$; however, this phase redefinition is not done in \texttt{SEOBNRv4HM}, and the corresponding expression reads $v_{\Omega }^5 \left(2 \pi + 12 i \log \left(2 \epsilon  v_{\Omega }\right)-17 i/3+12 i \gamma_E/3  \right)$. Therefore, when including the eccentric corrections in $T_{\ell m}^\text{ecc}$, we assume that the phase redefinition was done only for the eccentric part and keep using the same circular part as in \texttt{SEOBNRv4HM}.
In addition, since we expanded the tail part in eccentricity to $\Order(e^6)$, when factorizing the modes as in Eq.~\eqref{fEmodes} and writing the quasi-circular part in terms of frequency, we reexpand $T_{lm}^\text{ecc}$ in eccentricity (or $p_r$ and $\dot{p}_r$). For example, for the  $(2,2)$ mode, we obtain
\begin{equation}
T_{22}^\text{ecc} = -\frac{\pi }{4 r}\! \left[4 r^{3/2} \dot{p}_r+i p_r \left(r^2 \dot{p}_r+6\right)+2 \sqrt{r} p_r^2 + \Order(p_r^3)\right].
\end{equation}
The full expressions for $T_{lm}^\text{ecc}$ and  $f_{lm}^\text{ecc}$ are provided in the Supplemental Material~\cite{ancmaterial}.

\section{Conclusions}
\label{sec:conc}

Extending the waveform models used today in GW astronomy from quasi-circular to eccentric orbits is important for
future observations with LIGO, Virgo and KAGRA detectors~\cite{Abbott:2020qfu}, and with new facilities on the 
ground (Cosmic Explorer and Einstein Telescope), and in space (LISA). In fact, sources with non-negligible
eccentricity might come into reach of observations soon and should
routinely be included in searches and parameter inference.  While this
presents a challenge for waveform modeling and data analysis, it also
offers the unique opportunity to unveil the formation channels of compact binaries and probe
their environment (through eccentricity measurements).  In
this paper, we constructed an EOB waveform model for eccentric
binaries.  For this purpose, we obtained analytical results for the RR
force and waveform modes to 2PN order, including the leading-order
tail effects, and SO and SS couplings for aligned spins.

In particular, we first derived the RR force for eccentric orbits in PN expanded form, and then we recast it in a form that it can be directly incorporated 
in the quasi-circular RR force employed in the \texttt{SEOBNRv4HM}~\cite{Bohe:2016gbl,Cotesta:2018fcv} model, currently used in LIGO/Virgo analyses~\cite{LIGOScientific:2020ibl}.    
We then obtained initial conditions for the binary evolution which generalize those from Ref.~\cite{Buonanno:2005xu} to eccentric orbits, and which allow  
starting the binary's evolution from a specified initial frequency at periastron and an initial eccentricity (in the Keplerian parametrization). 
We also calculated all the waveform modes that contribute up to 2PN order relative to the leading order of the $(2,2)$ mode. It should be noted that the $(\ell,0)$ modes are proportional to the eccentricity and are hence important for eccentric orbits, especially the $(2,0)$ mode since it starts at the same PN order as the $(2,2)$ mode. Also the gravitational modes were 
rewritten in a factorized form to be straightforwardly implemented in the \texttt{SEOBNRv4HM} model.

Our results for the RR force and modes are valid for moderate to high eccentricities during the inspiral phase, since we do not use an eccentricity expansion except for the tail part, which is known analytically as an infinite series expansion. We provided expressions for the tail part in an expansion to $\Order(e^6)$, but we checked that expanding to $\Order(e^{10})$ produces negligible difference on the waveform even for high eccentricities ($\lesssim 0.9$). If results for $e$ close to 1 are needed, one could calculate the series expansion for the tail part numerically, or use analytical resummation methods as was done in Refs.~\cite{Loutrel:2016cdw,Tanay:2016zog}.

We are currently incorporating the eccentric RR force and gravitational modes of this paper in the inspiral-merger-ringdown 
quasi-circular--orbit \texttt{SEOBNRv4HM} waveform model (\texttt{SEOBNRv4EHM}~\cite{Buades:2021}) and validating it against NR simulations with eccentricity. We leave 
to future work the extension of the model to higher PN orders and the inclusion of spin precession.

\section*{Acknowledgments}
We are grateful to Serguei Ossokine, Harald Pfeiffer, Antoni Ramos-Buades, Hannes R\"{u}ter, and Maarten van de Meent for helpful discussions. We also thank Marco Stella, Marta Orselli, and Andrea Placidi for pointing out typos in Eq.~(A4).

\appendix

\section{Coordinate transformation from harmonic to EOB coordinates}
\label{app:transform}

The coordinate transformation from harmonic to EOB coordinates with no spin is given in Appendix~A of Ref.~\cite{Bini:2012ji}. In this appendix, we include LO SO and SS contributions to the transformation.
We label harmonic, ADM, and EOB coordinates by $(\bm{x}_h, \bm{v}_h)$, $(\bm{x}_a, \bm{p}_a)$, and $(\bm{x}, \bm{p})$, respectively.

\subsection{ADM to EOB transformation}
To find the canonical transformation from the ADM Hamiltonian with LO SO and SS using the NW SSC (see e.g. Refs.~\cite{damour1988higher,Damour:2000we,Steinhoff:2008zr}) and the 2PN expansion of the EOB Hamiltonian of Ref.~\cite{Barausse:2009xi}, we write an ansatz with unknown coefficients for the generating function $G(\bm{x}, \bm{p})$, perform the following transformation on the ADM Hamiltonian~\cite{Buonanno:1998gg}:
\begin{align}
x_a^i &= x^i + \frac{\partial G}{\partial p_i} - \frac{\partial G}{\partial x^j} \frac{\partial^2 G}{\partial p_j\partial p_i} + \Order\left(\frac{1}{c^6}\right), \nonumber\\
p_a^i &= p^i - \frac{\partial G}{\partial x_i} + \frac{\partial G}{\partial x^j} \frac{\partial^2 G}{\partial p_j\partial x^i} + \Order\left(\frac{1}{c^6}\right),
\end{align}
and match it to the EOB Hamiltonian to solve for the unknowns.

The result for the generating function is given by
\begin{widetext}
\begin{align}
G(x,p) &= \frac{p_r}{c^2} \left[-1-\frac{\nu }{2}+\frac{1}{2} \nu  p^2 r\right]
+ \frac{p_r}{c^4} \left[\frac{1}{8} \nu  (3 \nu -1) p^4 r-\frac{1}{8} \nu  (\nu +14) p^2-\frac{\nu ^2-7 \nu +1}{4 r}+\frac{1}{8} \nu ^2 p_r^2\right] \nonumber\\
&\quad +\frac{\nu^2}{2 c^4 r} \left[p_r (\hat{\bm{S}}_1 + \hat{\bm{S}}_2)^2 - (\bm{n}\cdot\hat{\bm{S}}_1 + \bm{n}\cdot\hat{\bm{S}}_2) (\bm{p}\cdot\hat{\bm{S}}_1 + \bm{p}\cdot\hat{\bm{S}}_2)\right],
\end{align}
which has no LO SO terms since the ADM and EOB Hamiltonian are the same at that order.
This generating function yields
\begin{align}
\label{ADMtoEOB}
\bm{x}_a &= \bm{x} + \frac{1}{c^2} \left[\bm{x} \left(\frac{\nu  p^2}{2}-\frac{\nu +2}{2 r}\right)+\nu  r p_r  \bm{p}\right] 
+ \frac{1}{c^4} \bigg\lbrace \bm{x} \left[\frac{3 (\nu -2) \nu  p^2}{8 r}-\frac{1}{8} \nu  (\nu +1) p^4-\frac{\nu  (5 \nu +16) p_r^2}{8 r}-\frac{\nu ^2-7 \nu +1}{4 r^2}\right] \nonumber\\
&\quad + \bm{p} p_r \left[\frac{1}{2} (\nu -1) \nu  p^2 r+\frac{(\nu -10) \nu}{4}\right]
+ \frac{\nu^2}{2r} \left[
(\hat{\bm{S}}_1 + \hat{\bm{S}}_2)^2 \frac{\bm{x}}{r}
 -  (\hat{\bm{S}}_1 + \hat{\bm{S}}_2) (\bm{n}\cdot\hat{\bm{S}}_1 + \bm{n}\cdot\hat{\bm{S}}_2) 
\right]
\bigg\rbrace, \nonumber\\
\bm{p}_a &= \bm{p} + \frac{1}{c^2} \left[\bm{p} \left(\frac{\nu +2}{2 r}-\frac{\nu  p^2}{2}\right)-\bm{x}\frac{(\nu +2)p_r}{2 r^2} \right]
+ \frac{1}{c^4} \bigg\lbrace
\bm{p} \left[\frac{1}{8} \nu  (3 \nu +1) p^4-\frac{\nu  (7 \nu +2) p^2}{8 r}+\frac{\nu  (\nu +8) p_r^2}{8 r}+\frac{2 \nu ^2-3 \nu +5}{4 r^2}\right] \nonumber\\
&\quad 
+ \bm{x} p_r \left[\frac{3 (\nu -2) \nu  p^2}{8 r^2}-\frac{3 \nu ^2-10 \nu +6}{4 r^3}+\frac{3 \nu ^2 p_r^2}{8 r^2}\right]
+ \frac{\nu^2 \bm{x}}{r^3} \left[(\hat{\bm{S}}_1 + \hat{\bm{S}}_2)^2 p_r - (\bm{n}\cdot\hat{\bm{S}}_1 + \bm{n}\cdot\hat{\bm{S}}_2) (\bm{p}\cdot\hat{\bm{S}}_1 + \bm{p}\cdot\hat{\bm{S}}_2)\right] \nonumber\\
&\quad
+ \frac{\nu^2}{2r^2} \left[-(\hat{\bm{S}}_1 + \hat{\bm{S}}_2)^2 \bm{p} + (\hat{\bm{S}}_1 + \hat{\bm{S}}_2) (\bm{p}\cdot\hat{\bm{S}}_1 + \bm{p}\cdot\hat{\bm{S}}_2)\right]
\bigg\rbrace.
\end{align}

\subsection{Harmonic to EOB transformation}

The transformation from harmonic to ADM coordinates is given by Eq.~(E1) of Ref.~\cite{Bini:2012ji}, which is independent of spin since the ADM and harmonic coordinates agree at LO SO and SS. Using that equation together with Eq.~\eqref{ADMtoEOB}, we obtain the following transformation from harmonic to EOB coordinates:
\begin{align}
\bm{x}_h &= \bm{x} + \frac{1}{c^2}  \left[\bm{x}\left(\frac{\nu  p^2}{2}-\frac{\nu +2}{2 r}\right)+\nu  r  p_r \bm{p}\right]
+ \frac{1}{c^4} \bigg\lbrace
\bm{x} \left[-\frac{1}{8} \nu  (\nu +1) p^4+\frac{(3\nu -1) \nu  p^2}{8 r}-\frac{\nu  (5 \nu +17) p_r^2}{8 r}-\frac{(\nu - 19)\nu}{4 r^2}\right] \nonumber\\
&\quad 
+ \bm{p} p_r \left[\frac{1}{4} (\nu -19) \nu +\frac{1}{2} (\nu -1) \nu p^2 r\right]
+ \frac{\nu^2}{2r} \left[
(\hat{\bm{S}}_1 + \hat{\bm{S}}_2)^2 \frac{\bm{x}}{r}
 -  (\hat{\bm{S}}_1 + \hat{\bm{S}}_2) (\bm{n}\cdot\hat{\bm{S}}_1 + \bm{n}\cdot\hat{\bm{S}}_2) 
\right]
\bigg\rbrace, \nonumber\\
\bm{v}_h &= \bm{p} + \frac{1}{c^2} \left[\bm{p} \left(\left(\nu -\frac{1}{2}\right) p^2-\frac{\nu +4}{2 r}\right)-\bm{x} \frac{(3 \nu +2)  p_r}{2 r^2}\right]
- \frac{1}{4c^3r^2} \left[\bm{n}\times\hat{\bm{S}}_1 (3 - 3\delta + 2 \nu) +\bm{n}\times\hat{\bm{S}}_2(3 + 3\delta + 2 \nu) \right] \nonumber\\
&\quad +\frac{1}{c^4} \bigg\lbrace
\bm{p} \left[\left(\frac{3}{8}-\nu \right) p^4+\frac{\left(7 \nu ^2-41 \nu +8\right) p^2}{8 r}+\frac{\left(-15 \nu ^2+29 \nu +8\right) p_r^2}{8 r}+\frac{-\nu ^2+15 \nu +1}{2 r^2}\right] \nonumber\\
&\quad
+ \bm{x} p_r \left[\frac{\left(4-7 \nu ^2-23 \nu\right) p^2}{8 r^2}+\frac{4-3 \nu ^2+9 \nu}{4 r^3}+\frac{3 \nu  (5 \nu +1) p_r^2}{8 r^2}\right] 
+ \frac{\nu^2}{2r^2} \left[ (\hat{\bm{S}}_1 + \hat{\bm{S}}_2) (\bm{p}\cdot\hat{\bm{S}}_1 + \bm{p}\cdot\hat{\bm{S}}_2)-(\hat{\bm{S}}_1 + \hat{\bm{S}}_2)^2 \bm{p}\right] \nonumber\\
&\quad
+ \frac{\nu^2 \bm{x}}{r^3} \left[(\hat{\bm{S}}_1 + \hat{\bm{S}}_2)^2 p_r - (\bm{n}\cdot\hat{\bm{S}}_1 + \bm{n}\cdot\hat{\bm{S}}_2) (\bm{p}\cdot\hat{\bm{S}}_1 + \bm{p}\cdot\hat{\bm{S}}_2)\right]
\bigg\rbrace,
\end{align}
and for the scalars ($\phi, r,\dot{\phi},\dot{r}$), we obtain
\begin{align}
\label{harmToEOB}
\phi_h &= \phi + \frac{\pphi \nu p_r}{c^2 r} + \frac{p_r\pphi}{c^4} \left[ \frac{3 (\nu -5) \nu }{4 r^2}-\frac{\nu  p^2}{2 r}-\frac{\nu ^2 p_r^2}{r}\right], \nonumber\\
r_h &= r + \frac{1}{c^2} \left(\frac{\nu}{2}   p^2 r+\nu  r p_r^2-1-\frac{\nu }{2}\right) +\frac{1}{c^4} \bigg[
\frac{\nu}{8}   (3 \nu -1) p^2-\frac{\nu}{8}   (\nu +1) p^4 r-\frac{\nu }{4 r}(\nu -19) +\frac{\nu}{2} (2 \nu -1) p^2 p_r^2 r\nonumber\\
&\quad\qquad
-\frac{\nu}{8} (3 \nu +55)p_r^2 -\frac{1}{2} \nu ^2 r p_r^4
+ \frac{1}{2r} \left(X_1^4 \chi_1^2 + 2 \nu^2 \chi_1\chi_2 + X_2^4 \chi_2^2\right)
\bigg] , \nonumber\\
\dot{\phi}_h &= \frac{\pphi}{r^2} + \frac{\pphi}{c^2r^2} \left[\frac{\nu -1 }{2}p^2-2 \nu p_r^2-\frac{1}{r}\right] 
+\frac{1}{2r^3c^3} \Big[\chi_1 (2+2 \delta -\nu)+\chi_2 (2-2 \delta -\nu )\Big]  +\frac{\pphi}{c^4r^2} \bigg[
4 \nu ^2 p_r^4
+\frac{\left(3 \nu ^2-17 \nu +2\right) p^2}{4 r}  \nonumber\\
&\quad\qquad
-2 (\nu -1) \nu  p_r^2 p^2
-\frac{\left(\nu ^2+5 \nu -3\right) p^4}{8} + \frac{4-5 \nu ^2+65 \nu}{4 r} p_r^2  -\frac{\nu ^2-9 \nu +2}{4 r^2}
+ \frac{1}{2r^2} \left(X_1^4 \chi_1^2 + 2 \nu^2 \chi_1\chi_2 + X_2^4 \chi_2^2\right)
\bigg], \nonumber\\
\dot{r}_h &= p_r  + \frac{p_r}{c^2} \left[ \left(2 \nu -\frac{1}{2}\right) p^2-\left(2 \nu +3\right)\frac{1}{r}-\nu  p_r^2\right]
+\frac{p_r}{c^4} \bigg[\left(\nu ^2-2 \nu +\frac{3}{8}\right) p^4+\frac{\left(\nu ^2-55 \nu +6\right) p^2}{4 r} + \left(\nu -\frac{5 \nu ^2}{2}\right) p_r^2p^2\nonumber\\
&\quad\qquad
+\frac{4-\nu ^2+39 \nu}{4 r}p_r^2+\frac{3}{2} \nu ^2 p_r^4+\frac{6-5 \nu ^2+39 \nu}{4 r^2}
+ \frac{1}{2r^2} \left(X_1^4 \chi_1^2 + 2 \nu^2 \chi_1\chi_2 + X_2^4 \chi_2^2\right)
\bigg].
\end{align}
\end{widetext}

\subsection{Transformation for the SSC}
When calculating the spin contributions to the waveform modes, we used the source moments from Refs.~\cite{Blanchet:2006gy,Buonanno:2012rv} which are in terms of the covariant SSC. To transform the resulting modes to the NW SSC, we use the center-of-mass shift~\cite{Kidder:1995zr}
\begin{equation}
{x_A^i}_\text{(cov)} \to x_A^i + \frac{1}{2c^3 m_A} (\bm{v}_A\times \bm{S}_A)^i,
\end{equation}
and the spin transformation \cite{Tagoshi:2000zg}
\begin{equation}
\bm{S}_1^\text{cov} = \left(1 - \frac{m_2}{c^2r}\right) \bm{S}_1 + \frac{1}{2c^2} \bm{v}_1 (\bm{v}_1\cdot \bm{S}_1),
\end{equation}
where the spin transformation is only required for the NLO SO part of the 2PN $(2,1)$ mode.

For the scalars ($r, \phi, \dot{r},\dot{\phi},\chi_1,\chi_2$), we obtain the transformations
\begin{align}
r_\text{cov} &= r - \frac{\nu r \dot{\phi}}{2 c^3} (\chi_1 + \chi_2), \nonumber\\
\phi_\text{cov} &= \phi + \frac{\nu \dot{r}}{2 c^3 r} (\chi_1 + \chi_2), \nonumber\\
\dot{r}_\text{cov} &= \dot{r} + \frac{\nu \dot{r}\dot{\phi}}{2c^3}(\chi_1 + \chi_2), \nonumber\\
\dot{\phi}_\text{cov} &= \dot{\phi} - \frac{\nu}{2 c^3 r^3} \left(1 + r \dot{r}^2 - r^3 \dot{\phi}^2\right) (\chi_1 + \chi_2) , \nonumber\\
\chi_1^\text{cov} &= \chi_1 - \frac{\chi_1}{2c^2 r} (1 - \delta), \nonumber\\
\chi_2^\text{cov} &= \chi_2 - \frac{\chi_2}{2c^2 r} (1 + \delta).
\end{align}

\section{Angular momentum flux at leading-order spin-squared}
\label{app:SSflux}
In this appendix, we derive the angular momentum flux at leading spin-squared (S$_i^2$) order.
Here, we use unscaled variables in \emph{harmonic} coordinates, but we drop the subscript `$h$' to simplify the notation. We denote the orbital angular momentum $\bm{L} = \mu \bm{r}\times \bm{v}$, the relative position $\bm{r} = \bm{x}_1 - \bm{x}_2$, and relative velocity $\bm{v} = d\bm{r}/dt$.

The relative acceleration $\bm{a} \equiv \bm{a}_1 - \bm{a}_2$ with LO SO and SS contributions, in harmonic coordinates and the NW SSC, is  given by~\cite{Kidder:1995zr}
\begin{align}
\bm{a} &= - M \frac{\bm{n}}{r^2} \nonumber\\
&\quad + 
\frac{1}{c^3} \bigg[
3\left(2 + \frac{3m_2}{2m_1}\right) \bm{n}\cdot(\bm{v}\times\bm{S}_1) \frac{\bm{n}}{r^3}
- \left(4 + \frac{3m_2}{m_1}\right)  \nonumber\\
&\quad\qquad
\frac{\bm{v}\times\bm{S}_1}{r^3}+ 3\left(2 + \frac{3m_2}{2m_1}\right)  \frac{\dot{r}}{r^3} \bm{n}\times\bm{S}_1
+ 1 \leftrightarrow 2
\bigg] \nonumber\\
&\quad -\frac{3}{c^4\mu r^4} \bigg[
\bm{n}  (\bm{S}_1\cdot\bm{S}_2) + \bm{S}_1 (\bm{n}\cdot\bm{S}_2) + \bm{S}_2(\bm{n}\cdot\bm{S}_1) \nonumber\\
&\quad\qquad - 5 \bm{n}(\bm{n}\cdot\bm{S}_1)(\bm{n}\cdot\bm{S}_2) \bigg]\nonumber\\
&\quad + \frac{3}{2c^4 r^4} \bigg[
\frac{m_2 \CESA}{m_1\mu} \bigg(-\bm{n} \, \bm{S}_1^2 + 5 \bm{n} (\bm{n}\cdot\bm{S}_1)^2 \nonumber\\
&\quad\qquad 
- 2 \bm{S}_1 (\bm{n}\cdot \bm{S}_1) \bigg)
+ 1\leftrightarrow 2 \bigg].
\end{align}
Since the spin evolution equations start at 1PN order, we can assume $\dot{\bm{S}}_1 = 0 = \dot{\bm{S}}_2$ for the calculation of the LO fluxes.

The source multipole moments needed are the spin quadrupole $I^{ij}$ and the current quadrupole $J^{ij}$, which are given by \cite{Kidder:1995zr,Blanchet:2006gy,Maia:2017yok}
\begin{align}
I^{ij} &= m_1 x_1^{\langle i} x_1^{j \rangle} + \frac{3}{c^3} x_1^{\langle i} (\bm{v}_1\times \bm{S}_1)^{j \rangle} - \frac{4}{3c^3} \frac{d}{dt} x_1^{\langle i}(\bm{x}_1\times \bm{S}_1)^{j \rangle} \nonumber\\
&\quad - \frac{\CESA}{c^4m_1} S_1^{\langle i} S_1^{j \rangle}  + 1 \leftrightarrow 2 \\
J^{ij} &= m_1 x_1^{\langle i} (\bm{x}_1 \times \bm{v}_1)^{j \rangle} + \frac{3}{2c} x_1^{\langle i} S_1^{j \rangle} + 1 \leftrightarrow 2,
\end{align}
where the indices in angle brackets denote a symmetric trace-free part.

To transform from the coordinates of the two bodies $x_1^i$ and $x_2^i$ to the center-of-mass relative coordinates $x^i = x_1^i - x_2^i$, we use \cite{Will:2005sn}
\begin{equation}
x_1^i = \frac{m_2}{M} x^i + \delta x^i, \quad x_2^i = - \frac{m_1}{M} x^i + \delta x^i,
\end{equation}
where
\begin{equation}
\delta x^i = -\frac{\nu}{2c^3} \left[\frac{(\bm{v} \times \bm{S}_1)^i}{m_1} - \frac{(\bm{v} \times \bm{S}_2)^i}{m_2} \right].
\end{equation}

The energy and angular momentum fluxes in terms of the multipole moments, to the order needed for the LO fluxes, are then calculated from \cite{Thorne:1980ru,Kidder:1995zr}
\begin{align}
\Phi_E &= \frac{1}{5} I_{ij}^{(3)} I_{ij}^{(3)} + \frac{16}{45c^2} J_{ij}^{(3)} J_{ij}^{(3)}, \\
\Phi_J^i &= \frac{2}{5} \epsilon_{ijk} I_{jl}^{(2)} I_{kl}^{(3)} + \frac{32}{45c^2} \epsilon_{ijk} J_{jl}^{(2)} J_{kl}^{(3)}.
\end{align}
This yields the LO SO and $\text{S}_1\text{S}_2$ fluxes derived in Refs.~\cite{Kidder:1995zr,Zeng:2007bq,Wang:2007ntb}, in addition to the $\text{S}_i^2$ energy flux from Ref.~\cite{Maia:2017yok}. For the $\text{S}_i^2$ angular momentum flux, we obtain
\begin{align}
\bm{\Phi}_J^{S_i^2} &= \frac{2m_2^2}{5c^4 r^5} \bigg[
\frac{\bm{L}}{\mu r} \bm{S}_1^2 - \bm{n}\cdot\bm{S}_1 (\bm{v} \times \bm{S}_1) + \bm{v}\cdot\bm{S}_1 (\bm{n} \times \bm{S}_1)
\bigg] \nonumber\\
&\quad + \frac{2m_2^2 \CESA}{5c^4 M r^4} \bigg[
\frac{\bm{L}}{\mu r} \bm{S}_1^2 \left(-30 \dot{r}^2 + 12 v^2 + 24 \frac{M}{r}\right) \nonumber\\
&\quad\qquad 
+\frac{\bm{L}}{\mu r} (\bm{n}\cdot\bm{S}_1)^2 \left(210 \dot{r}^2 - 60 v^2 - 90 \frac{M}{r}\right) \nonumber\\
&\quad\qquad 
+\bm{v}\times \bm{S}_1 (\bm{n}\cdot\bm{S}_1) \left(30 \dot{r}^2 - 18 v^2 - 12 \frac{M}{r}\right)\nonumber\\
&\quad\qquad 
+ 6 \bm{n}\times \bm{S}_1 (\bm{v}\cdot\bm{S}_1 - \dot{r} \bm{n}\cdot\bm{S}_1) \frac{M}{r} \nonumber\\
&\quad\qquad 
-90 \frac{\bm{L}}{\mu r} \dot{r} (\bm{n}\cdot\bm{S}_1) (\bm{v}\cdot\bm{S}_1) 
+ 6 \frac{\bm{L}}{\mu r} \left(\bm{v}\cdot\bm{S}_1\right)^2 \nonumber\\
&\quad\qquad 
- 6 \dot{r} (\bm{v}\cdot\bm{S}_1) \bm{v}\times\bm{S}_1
\bigg] + 1\leftrightarrow 2.
\end{align}
This is in agreement with the recent results of Ref.~\cite{Cho:2021mqw}, although our expression appears simpler because of using the individual spins $S_i$ and masses $m_i$, instead of different combinations of them.

\section{Aligned-spin contributions to the modes in harmonic coordinates}
\label{app:harmModes}

The modes calculated from the source moments of Refs.~\cite{Blanchet:2006gy,Buonanno:2012rv} in \emph{harmonic} coordinates and using the \emph{covariant} SSC have the following spin contributions:
\begin{widetext}
\begin{align}
\hat{H}_\text{spin}^{20} &= \frac{\dot{\phi}}{\sqrt{6} c^3 r} \left[\chi_1 (1 + \delta +\nu) + \chi_2 (1 - \delta +\nu)\right]
- \frac{\sqrt{3} }{2\sqrt{2} c^4 r^3} \left[\CESA X_1^2 \chi_1^2 + 2 \nu \chi_1 \chi_2 + \CESB X_2^2 \chi_2^2\right], \nonumber\\
\hat{H}_\text{spin}^{21} &= - \frac{i}{4r^2} \left[(1 + \delta )\chi_1 + (\delta - 1)\chi_2 \right] + \frac{i}{168 c^4 r^3} \bigg\lbrace
\chi_1 \bigg[154+22 \delta  (\nu +7)+34 \nu +4 r^3 \dot{\phi }^2 (4 \nu\delta -21\delta+66 \nu -21) \nonumber\\
&\quad
-2 i \dot{r} r^2 \dot{\phi } (13 \nu\delta +147\delta-83 \nu +147)+\dot{r}^2 r (-60 \delta  \nu +105 \delta -52 \nu +105)\bigg]
+\chi_2 \bigg[-154 + 22 \delta  (\nu +7)-34 \nu \nonumber\\
&\quad
+4 r^3 \dot{\phi }^2 (4 \nu\delta -21\delta -66 \nu +21)-2 i \dot{r} r^2 \dot{\phi } (13 \nu\delta +147\delta+83 \nu -147) +\dot{r}^2 r (-60 \delta  \nu +105 \delta +52 \nu -105)\bigg]
\bigg\rbrace, \nonumber\\
\hat{H}_\text{spin}^{22} &= -\frac{1}{6c^3 r^2} \left\lbrace
\chi _1 \left[r \dot{\phi } (3 \delta -5 \nu +3)+i \dot{r} (3 \delta -8 \nu +3)\right]
+\chi _2 \left[r \dot{\phi } (-3 \delta -5 \nu +3)-i \dot{r} (3 \delta +8 \nu -3)\right]
\right\rbrace \nonumber\\
&\quad
+ \frac{3}{4c^4 r^3} \left[\CESA \chi _1^2 X_1^2+\CESB \chi _2^2 X_2^2+2 \nu  \chi _2 \chi _1\right], \nonumber\\
\hat{H}_\text{spin}^{30} &= -\frac{i \nu  \dot{r}}{\sqrt{42}c^3 r^2} \left(\chi _1+\chi _2\right), \nonumber\\
\hat{H}_\text{spin}^{31} &= \frac{i}{48\sqrt{14} c^4 r^3} \bigg\lbrace
\chi_1 \bigg[
-4 + 20 \delta  \nu -4 \delta +20 \nu +r^3 \dot{\phi }^2 (-31 \delta  \nu -24 \delta +87 \nu -24)+i\dot{r} r^2 \dot{\phi } (-70 \delta  \nu -12 \delta +62 \nu -12) \nonumber\\
&\quad\quad 
+\dot{r}^2 r (-30 \delta  \nu +12 \delta -50 \nu +12)\bigg]
+ \chi_2 \bigg[
4+20 \delta  \nu -4 \delta -20 \nu +r^3 \dot{\phi }^2 (-31 \delta  \nu -24 \delta -87 \nu +24) \nonumber\\
&\quad\quad
+i\dot{r} r^2 \dot{\phi } (-70 \delta  \nu -12 \delta -62 \nu +12)+\dot{r}^2 r (-30 \delta  \nu +12 \delta +50 \nu -12)
\bigg]
\bigg\rbrace, \nonumber\\
\hat{H}_\text{spin}^{32} &= \sqrt{\frac{5}{7}} \frac{ \nu }{6c^3 r^2} \left(4 r \dot{\phi }+i \dot{r}\right) \left(\chi _1+\chi _2\right), \nonumber\\
\hat{H}_\text{spin}^{33} &= \sqrt{\frac{5}{42}} \frac{i}{16 c^4 r^3} \bigg\lbrace
\chi_1 \bigg[
4-20 \delta  \nu +4 \delta -20 \nu +r^3 \dot{\phi }^2 (-33 \delta  \nu +24 \delta -119 \nu +24)+i\dot{r} r^2 \dot{\phi } (-78\delta  \nu +36 \delta -154 \nu +36) \nonumber\\
&\quad\quad
+\dot{r}^2 r (30 \delta  \nu -12 \delta +50 \nu -12)\bigg]
+\chi_2 \bigg[
-4-20 \delta  \nu +4 \delta +20 \nu +r^3 \dot{\phi }^2 (-33 \delta  \nu +24 \delta +119 \nu -24) \nonumber\\
&\quad\quad
+i\dot{r} r^2 \dot{\phi } (-78\delta  \nu +36\delta +154 \nu -36)+\dot{r}^2 r (30 \delta  \nu -12 \delta -50 \nu +12)
\bigg]
\bigg\rbrace, \nonumber\\
\hat{H}_\text{spin}^{41} &=  -i \sqrt{\frac{5}{2}} \frac{\nu}{336 c^4 r^3} \left(11 r^3 \dot{\phi }^2-10 i \dot{r} r^2 \dot{\phi }+6 \dot{r}^2 r-12\right) \left[(\delta -1) \chi _1+(\delta +1) \chi _2\right], \nonumber\\
\hat{H}_\text{spin}^{43} &= \sqrt{\frac{5}{14}} \frac{\nu}{48 c^4 r^3}  \left(-23 i r^3 \dot{\phi }^2+10 \dot{r} r^2 \dot{\phi }+2 i \dot{r}^2 r-4 i\right) \left[(\delta -1) \chi _1+(\delta +1) \chi _2\right].
\end{align}
\end{widetext}

\section{Keplerian parametrization}
\label{app:kepler}
This appendix provides expressions for some orbital quantities in the Keplerian parametrization that are needed for calculating the initial conditions, and the tail part of the RR force and waveform modes.

In the Keplerian parametrization,
\begin{equation}
r = \frac{1}{u_p (1 + e \cos \chi)},
\end{equation} 
where $u_p$ is the inverse semilatus rectum and $\chi$ is the relativistic anomaly.
Inverting the Hamiltonian at the turning points $r_\pm = 1 / (u_p(1 \pm e))$ and solving for the energy and angular momentum to 2PN order yields
\begin{align}
\label{ELeup}
E &= \frac{1}{2} \left(e^2-1\right) u_p -\frac{u_p^2}{8c^2} \left(e^2-1\right)^2 (\nu -3)  \nonumber\\
&\quad +\frac{u_p^3}{16 c^4} \left(e^2-1\right)^2 \left[e^2 \left(\nu ^2-3 \nu +5\right)-\nu ^2-5 \nu +27\right] \nonumber\\
&\quad + \frac{\left(1-e^2\right)^2 u_p^{5/2}}{4c^4}  \left[\chi _1 (\nu-2 \delta -2)+\chi _2 (\nu + 2 \delta -2)\right] \nonumber\\
&\quad +\frac{\left(1-e^2\right)^2 u_p^3}{4c^4}  \Big[\chi _1^2 \left(\CESA X_1^2 + X_1^4\right) + \nonumber\\
&\quad\qquad +2 \nu (1 + \nu)  \chi _2 \chi _1 + \chi _2^2 \left(\CESB X_2^2 + X_2^4\right)\Big], \nonumber\\
\pphi &= \frac{1}{\sqrt{u_p}} + \frac{\sqrt{u_p}}{2c^2} \left(e^2+3\right)  +\frac{u_p^{3/2}}{8c^4} \left(e^2+3\right)  \left(3 e^2-4 \nu +9\right) \nonumber\\
&\quad +\frac{1}{4} \left(e^2+3\right) u_p \left[\chi _1 (-2 \delta +\nu -2)+\chi _2 (2 \delta +\nu -2)\right] \nonumber\\
&\quad +  \frac{\left(e^2+3\right) u_p^{3/2}}{4c^4}  \Big[\chi _1^2 \left(\CESA X_1^2 + X_1^4\right) \nonumber\\
&\quad\qquad +\chi _2^2 \left(\CESB X_2^2 + X_2^4\right)\Big] \nonumber\\
&\quad + \frac{\nu u_p^{3/2} \chi _1 \chi _2 }{2 c^4} \left[e^2 (3 \nu +1)+\nu +3\right].
\end{align}
Inverting $\pphi(u_p,e)$, we obtain $u_p(\pphi,e)$
\begin{align}
\label{upLe}
u_p(\pphi,e) &= \frac{1}{\pphi^2}+\frac{\left(e^2+3\right)}{c^2 \pphi^4} + \frac{\left(e^2+3\right)\left(2 e^2-\nu +6\right)}{c^4 \pphi^6} \nonumber\\
&\quad +\frac{3+e^2}{2 \pphi^5c^3} \left[
  (\nu-2 \delta -2) \chi_1 +  (\nu+2 \delta -2) \chi_2
\right] \nonumber\\
&\quad + \frac{1}{2 \pphi^6 c^4} \bigg\lbrace 
 \nu  \chi _1 \chi _2 \left[e^2 (3 \nu +1)+\nu +3\right] \nonumber\\
&\quad\quad+ \chi_1^2 \left[\CESA \left(e^2+3\right) X_1^2 + \left(3 e^2+1\right) X_1^4\right] \nonumber\\
&\quad\quad + 1 \leftrightarrow 2\bigg\rbrace.
\end{align}
Inverting the Hamiltonian to obtain $p_r(E,\pphi)$, and plugging $E(e,u_p)$ and $\pphi(e,u_p)$, yields
\begin{align}
p_r &= e \sqrt{u_p} \sin\chi 
+\frac{e u_p^{3/2}}{2c^2}  \sin\chi \left(e^2+2 e \cos\chi+1\right) +\dots
\end{align}

The radial and azimuthal periods are given, respectively, by
\begin{align}
T_r &= \oint dt = \oint \left(\frac{\partial H}{\partial p_r}\right)^{-1} dr = 2\int_{0}^{\pi}  \left(\frac{\partial H}{\partial p_r}\right)^{-1} \frac{dr}{d\chi} d\chi \nonumber\\
& = \frac{2 \pi }{\left(u_p-e^2 u_p\right)^{3/2}} -\frac{\pi  (\nu -6)}{c^2\sqrt{u_p-e^2 u_p}} + \dots, \nonumber\\
T_\phi &=  \oint \dot{\phi} dt = \oint \frac{\partial H}{\partial \pphi} dt = 2 \pi + \frac{6 \pi  u_p}{c^2} + \dots.
\end{align}
The associated frequencies are
\begin{equation}
\omega_r = \frac{2\pi}{T_r}, \qquad
\omega_\phi = \frac{T_\phi}{T_r}.
\end{equation}
The dimensionless frequency variable $x \equiv \omega_\phi^{2/3}$ to 2PN order is given by
\begin{align}
x &= u_p-e^2 u_p + \frac{u_p^2}{3c^2} \left(e^2-1\right) \left(e^2 (\nu -6)-\nu \right) \nonumber\\
&\quad -\frac{u_p^3}{36c^4} \left(e^2-1\right)\bigg\lbrace
e^4 \left(8 \nu ^2-33 \nu +180\right) +8 \nu ^2 \nonumber\\
&\quad
-2 e^2 \left[3 \left(12 \sqrt{1-e^2}+5\right) \nu -90 \sqrt{1-e^2}+8 \nu ^2+27\right] \nonumber\\
&\quad +9 \left(8 \sqrt{1-e^2}-13\right) \nu -180 \left(\sqrt{1-e^2}-1\right)
\bigg\rbrace \nonumber\\
&\quad + \frac{ u_p^{5/2}}{6c^4}\left(e^2-1\right) \left(3 e^2+1\right) \left[\chi_1 (2+2 \delta -\nu) + 1\leftrightarrow 2\right] \nonumber\\
&\quad -\frac{u_p^3}{2c^4} \Big\lbrace \chi_1^2 \left[\CESA \left(e^4-1\right) X_1^2 + \left(e^2-1\right)^2 X_1^4 \right] \nonumber\\
&\quad\quad + \nu  \chi _1 \chi _2 \left[e^4 (\nu +1)-2 e^2 \nu +\nu -1\right]
+ 1\leftrightarrow 2
\Big\rbrace,
\end{align}
which can be inverted to obtain $u_p(x,e)$, the 1PN part of which reads
\begin{align}
\label{upxe}
u_p &= -\frac{x}{e^2-1} + \frac{x^2 \left[e^2 (\nu -6)-\nu \right]}{3c^2 \left(e^2-1\right)^2} + \dots.
\end{align}

\section{$\dot{p}_r$ in tortoise coordinates}
\label{app:tort}
The tortoise-coordinate $r_*$ is defined by~\cite{Damour:2007xr,Pan:2009wj}
\begin{equation}
\frac{dr_*}{dr} = \frac{\sqrt{D(r)}}{A(r)} \equiv \frac{1}{\xi(r)},
\end{equation}
where $A(r)$ and $D(r)$ are the metric potentials
\begin{equation}
ds^2_\text{eff} = - A(r) dt^2 + \frac{D(r)}{A(r)} dr^2 + r^2 d\Omega^2. 
\end{equation}
The conjugate momentum to $r_*$ is denoted $p_{r_*}$, and invariance of the action gives the relation
\begin{equation}
\label{prstar}
p_{r_*} =  p_r \xi(r).
\end{equation}

The Hamiltonian and EOMs used in {SEOBNRv4} (see Eqs.~(10) of Ref.~\cite{Pan:2011gk}) are expressed in terms of the variables $(r,p_{r_*},\phi,\pphi)$. However, the RR force we derived in Sec.~\ref{sec:RRforce} is expressed in terms of $(r,p_r,\dot{p}_r)$. We use Eq.~\eqref{prstar} to replace $p_r$ with $p_{r_*}$, and to obtain a relation between $\dot{p}_r$ and the derivatives of $H_\text{EOB}(r,p_{r_*},p_\phi)$.
We use the following relations:
\begin{align}
dH &= \left(\frac{\partial H}{\partial r}\right)_{p_{r_*}} \! dr + \left(\frac{\partial H}{\partial p_{r_*}}\right)_{r} \! dp_{r_*} + \frac{\partial H}{\partial p_\phi} dp_\phi \nonumber\\
 & = \left(\frac{\partial H}{\partial r}\right)_{p_r} \! dr + \left(\frac{\partial H}{\partial p_r}\right)_{r} \! dp_r + \frac{\partial H}{\partial p_\phi} dp_\phi, \\
dp_{r_*} &= \left(\frac{\partial p_{r_*}}{\partial r}\right)_{p_r} dr + \left(\frac{\partial p_{r_*}}{\partial r}\right)_r dp_r,
\end{align}
leading to 
\begin{align}
\left(\frac{\partial H}{\partial r}\right)_{p_r} &= \left(\frac{\partial H}{\partial r}\right)_{p_{r_*}} + \left(\frac{\partial H}{\partial p_{r_*}}\right)_{r} \left(\frac{\partial p_{r_*}}{\partial r}\right)_{p_r},
\end{align}
where
\begin{equation}
\left(\frac{\partial p_{r_*}}{\partial r}\right)_{p_r} = p_r \frac{d \xi(r)}{d r}.
\end{equation}
Hence,
\begin{align}
\label{prdotstar}
\dot{p}_r &= - \left(\frac{\partial H}{\partial r}\right)_{p_r} \nonumber\\
&= - \left[\left(\frac{\partial H}{\partial r}\right)_{p_{r_*}} + \left(\frac{\partial H}{\partial p_{r_*}}\right)_{r} \frac{p_{r_*}}{\xi(r)} \frac{d \xi(r)}{d r} \right].
\end{align}

\bibliography{references}


\end{document}